\documentclass[a4paper,fleqn,usenatbib]{mnras}

\usepackage[T1]{fontenc}
\usepackage{ae,aecompl}

\usepackage{graphicx,subfig}	% Including figure files
\usepackage{amsmath}	% Advanced maths commands
\usepackage{amssymb}	% Extra maths symbols 
\usepackage[varg]{txfonts}

\usepackage[scr=rsfs, scrscaled=1.05,cal=euler]{mathalfa}

\DeclareMathAlphabet{\mathpzc}{OT1}{pzc}{bx}{it}

\numberwithin{equation}{section}

\newcommand{\EllF}{\mathrm{F}}
\newcommand{\EllK}{\mathrm{K}}  
\newcommand{\ka}{\kappa_1}
\newcommand{\kb}{\kappa_2}

\newcommand{\tka}{\tilde{\kappa}_1}
\newcommand{\tkb}{\tilde{\kappa}_2}

\newcommand{\E}{\mathcal{E}}  
\newcommand{\oder}[2]{\dfrac{\mathrm{d}{#1}}{\mathrm{d}{#2}}}
\newcommand{\pder}[2]{\dfrac{\partial{#1}}{\partial{#2}}}

\DeclareMathOperator{\arctanh}{arctanh}
\DeclareMathOperator{\arccosh}{arccosh}

\DeclareMathOperator{\sn}{sn}
\DeclareMathOperator{\cn}{cn}
\DeclareMathOperator{\dn}{dn}

\DeclareMathOperator{\cd}{cd}
\DeclareMathOperator{\dc}{dc}

\DeclareMathOperator{\sign}{sign}
\newcounter{subeqn} %
\makeatletter
\@addtoreset{subeqn}{equation}
\makeatother

\title[Nonconservative extension of Keplerian integrals]{Nonconservative extension of Keplerian integrals and a new class of integrable system}

\author[J. Roa]{
Javier Roa\thanks{\emph{Present address:} Jet Propulsion Laboratory, California Institute of Technology, 4800 Oak Grove Drive, Pasadena, CA 91109-8099, USA. E-mail: javier.roa@upm.es (JR)}
\\
Space Dynamics Group, Technical University of Madrid, Pza. Cardenal Cisneros 3, Madrid, E-28040, Spain
}

\date{Accepted 2016 August 31. Received 2016 August 25; in original form 2016 June 21\\doi:\href{https://dx.doi.org/10.1093/mnras/stw2209}{10.1093/mnras/stw2209}\vspace{-3mm}}

\pubyear{2016}

\begin{document}
\label{firstpage}
\pagerange{\pageref{firstpage}--\pageref{lastpage}}
\maketitle

\begin{abstract}
The invariance of the Lagrangian under time translations and rotations in Kepler's problem yields the conservation laws related to the energy and angular momentum. Noether's theorem reveals that these same symmetries furnish generalized forms of the first integrals in a special nonconservative case, which approximates various physical models. The system is perturbed by a biparametric acceleration with components along the tangential and normal directions. A similarity transformation reduces the biparametric disturbance to a simpler uniparametric forcing along the velocity vector. The solvability conditions of this new problem are discussed, and closed-form solutions for the integrable cases are provided. Thanks to the conservation of a generalized energy, the orbits are classified as elliptic, parabolic, and hyperbolic. Keplerian orbits appear naturally as particular solutions to the problem. After characterizing the orbits independently, a unified form of the solution is built based on the Weierstrass elliptic functions. The new trajectories involve fundamental curves such as cardioids and logarithmic, sinusoidal, and Cotes' spirals. These orbits can represent the motion of particles perturbed by solar radiation pressure, of spacecraft with continuous thrust propulsion, and some instances of Schwarzschild geodesics. Finally, the problem is connected with other known integrable systems in celestial mechanics.
\end{abstract}


%%%%%%%%%%%%%%%%%%%%%%%%%%%%%%%%%%%%%%%%%%%%%%%%%%

%%%%%%%%%%%%%%%%% BODY OF PAPER %%%%%%%%%%%%%%%%%%

\section{Introduction}
Finding first integrals is fundamental for characterizing a dynamical system. The motion is confined to submanifolds of lower dimensions on which the orbits evolve, providing an intuitive interpretation of the dynamics and reducing the complexity of the system. In addition, conserved quantities are good candidates when applying the second method of Lyapunov for stability analysis. Conservative systems related to central forces are typical examples of (Liouville) integrability, and provide useful analytic results. Hamiltonian systems have been widely analyzed in the classical and modern literature to determine adequate integrability conditions. The existence of first integrals under the action of small perturbations occupied \citet[][Chap.~V]{poincare1892methodes} back in the 19th century. Later, Emmy \citet{noether1918invariante} established in her celebrated theorem that conservation laws can be understood as the system exhibiting dynamical symmetries. In a more general framework, \citet{yoshida1983necessary,yoshida1983necessaryII} analyzed the conditions that yield algebraic first integrals of generic systems. He relied on the Kowalevski exponents for characterizing the singularities of the solutions and derived the necessary conditions for existence of first integrals exploiting similarity transformations.

Conservation laws are sensitive to perturbations and their generalization is not straightforward. For example, the Jacobi integral no longer holds when transforming the circular restricted three-body problem to the elliptic case \citep{xia1993arnold}. Nevertheless, \citet{contopoulos1967integrals} was able to find approximate conservation laws for orbits of small eccentricities. \citet{szebehely1964elliptic} benefited from the similarities between the elliptic and the circular problems in order to define transformations connecting them. \citet{henon1964applicability} deepened in the nature of conservation laws and reviewed the concepts of isolating and nonisolating integrals. Their study introduced a similarity transformation that embeds one of the constants of motion and transforms the original problem into a simplified one, reducing the degrees of freedom \citep[][\S3.2]{arnold2007mathematical}. \citet{carpintero2008finding} proposed a numerical method for finding the dimension of the manifold in which orbits evolve, i.e. the number of isolating integrals that the system admits.

The conditions for existence of integrals of motion under nonconservative perturbations received important attention in the past due to their profound implications. \citet{djukic1975noether} advanced on Noether's theorem and included nonconservative forces in the derivation. Relying on Hamilton's variational principle, they not only extended Noether's theorem, but also its inverse form and the Noether-Bessel-Hagen and Killing equations. Later studies by \citet{djukic1984integrating} sought integrating factors that yield conservation laws upon integration. Examples of application of Noether's theorem to constrained nonconservative systems can be found in the work of \citet{bahar1987extension}.  \citet{honein1991conservation} arrived to a compact formulation using what was later called the neutral action method. Remarkable applications exploiting Noether's symmetries span from cosmology \citep{capozziello2009noether,basilakos2011using} to string theory \citep{beisert2008dual}, field theory \citep{halpern1977field}, and fluid models \citep{narayan1987physics}. In the book by \citet[Chaps.~4 and 5]{olver2000applications}, an exhaustive review of the connection between symmetries and conservation laws is provided within the framework of Lie algebras. We refer to \citet[][Chap.~3]{arnold2007mathematical} for a formal derivation of Noether's theorem, and a discussion on the connection between conservation laws and dynamical symmetries.

%\newpage

Integrals of motion are often useful for finding analytic or semi-analytic solutions to a given problem. The acclaimed solution to the satellite main problem by \citet{brouwer1959solution} is a clear example of the decisive role of conserved quantities in deriving solutions in closed form. By perturbing the Delaunay elements, \citet{brouwer1961theoretical} solved the dynamics of a satellite subject to atmospheric drag and the oblateness of the primary. They proved the usefulness of canonical transformations even in the context of nonconservative problems. \citet[][pp.~81--82]{whittaker1917treatise} approached the problem of a central force depending on powers of the radial distance, $r^n$, and found that there are only fourteen values of $n$ for which the problem can be integrated in closed form using elementary functions or elliptic integrals. Later, he discussed the solvability conditions for equations involving square roots of polynomials \citep[p.~512]{whittaker1927course}. \citet{broucke1980notes} advanced on Whittaker's results and found six potentials that are a generalization of the integrable central forces discussed by the latter. These potentials include the referred fourteen values of $n$ as particular cases. Numerical techniques for shaping the potential given the orbit solution were published by \citet{carpintero1998orbit}. Classical studies on the integrability of systems governed by central forces are based strongly on Newton's theorem of revolving orbits.\footnote{Section IX, Book I, of Newton's Principia is devoted to the motion of bodies in moveable orbits (\emph{De Motu Corporum in Orbibus mobilibus, deq; motu Apsidum}, in the original latin version). In particular, Thm.~XIV states that ``The difference of the forces, by which two bodies may be made to move equally, one in a quiescent, the other in the same orbit revolving, is in a triplicate ratio of their common altitudes inversely''. Newton proved this theorem relying on elegant geometric constructions. The motivation behind this result was the development of a theory for explaining the precession of the orbit of the Moon. A detailed discussion about this theorem can be found in the book by \citet[pp.~184--201]{chandrasekhar1995newton}} The problem of the orbital precession caused by central forces was recently recovered by \citet{adkins2007orbital}, who considered potentials involving both powers and logarithms of the radial distance, and the special case of the Yukawa potential \citep{yukawa1935interaction}. \citet{chashchina2008remark} relied on Hamilton's vector to simplify the analytic solutions found by \citet{adkins2007orbital}. More elaborated potentials have been explored for modeling the perihelion precession \citep{schmidt2008perihelion}. The dynamics of a particle in Schwarzschild space-time can also be regarded as orbital motion perturbed by an effective potential depending on inverse powers of the radial distance \citep[][p.~102]{chandrasekhar1983mathematical}.

Potentials depending linearly on the radial distance appear recursively in the literature because they render constant radial accelerations, relevant for the design of spacecraft trajectories propelled by continuous-thrust systems. The pioneering work by \citet{tsien1953take} provided the explicit solution to the problem in terms of elliptic integrals, as predicted by \citet[][p.~81]{whittaker1917treatise}. By means of a special change of variables, \citet{izzo2015explicit} arrived to an elegant solution in terms of the Weierstrass elliptic functions. These functions were also exploited by \citet{macmillan1908motion} when he solved the dynamics of a particle attracted by a central force decreasing with $r^{-5}$. \citet{Urrutxua2015} solved the Tsien problem using the Dromo formulation, which models orbital motion with a regular set of elements \citep{pelaez2007special,urrutxua2015dromo,roa2015singularities}. Advances on Dromo can be found in the works by \citet{bau2015non} and \citet{roa2015orbit}. The case of a constant radial force was approached by \citet{akella2002anatomy} from an energy-driven perspective. They studied in detail the roots of the polynomial appearing in the denominator of the equation to integrate, and connected their nature with the form of the solution. General considerations on the integrability of the problem can be found in the work of \citet{san2012bounded}. 

Another relevant example of an integrable system in celestial mechanics is the Stark problem, governed by a constant acceleration fixed in the inertial frame. \citet{lantoine2011complete} provided the complete solution to the motion relying extensively on elliptic integrals and Jacobi elliptic functions. A compact form of the solution involving the Weierstrass elliptic functions was later presented by \citet{biscani2014stark}, who also exploited this formalism for building a secular theory for the post-Newtonian model \citep{biscani2012first}. The Stark problem provides a simplified model of radiation pressure. In the more general case, the dynamics subject to this perturbation cannot be solved in closed form. An intuitive simplification that makes the problem integrable consists in assuming that the force due to the solar radiation pressure follows the direction of the Sun vector. The dynamics are equivalent to those governed by a Keplerian potential with a modified gravitational parameter. 

The present paper introduces a new class of integrable system, governed by a biparametric nonconservative perturbation. This acceleration unifies various force models, including special cases of solar radiation pressure, low-thrust propulsion, and some particular configurations in general relativity. The problem is formulated in Sec.~\ref{Sec:Dynamics}, where the biparametric acceleration is defined and then reduced to a uniparametric forcing thanks to a similarity transformation. The conservation laws for the energy and angular momentum are generalized to the nonconservative case by exploiting known symmetries of Kepler's problem. Before solving the dynamics explicitly, we will prove that there are four cases that can be solved in closed form using elementary or elliptic functions. Sections~\ref{Sec:conic}--\ref{Sec:sinusoidal} present the properties of each family of orbits and the corresponding trajectories are derived analytically. Section~\ref{Sec:summary} is a summary of the solutions, which are unified in Sec.~\ref{Sec:Weierstrass} introducing the Weierstrass elliptic functions. Finally, Sec.~\ref{Sec:connection} discusses the connection with known solutions to similar problems, and with Schwarzschild geodesics.

\section{Dynamics}\label{Sec:Dynamics}
The motion of a body orbiting a central mass under the action of an arbitrary perturbation $\mathbfit{a}_p$ is governed by
\begin{equation}\label{Eq:two_body_problem}
	\oder{^2\mathbfit{r}}{t^2}  =-\frac{\mu}{r^3}\mathbfit{r} +\mathbfit{a}_p,
\end{equation}
where $\mu$ denotes the gravitational parameter of the attracting body, $\mathbfit{r}$ is the radiusvector of the particle, and $r=||\mathbfit{r}||$. In the case $\mathbfit{a}_p=0$, Eq.~\eqref{Eq:two_body_problem} reduces to Kepler's problem, which can be solved in closed form. 

It is well known that under the action of a radial perturbation of the form
\begin{equation}
	\mathbfit{a}_p = \frac{\xi\mu}{r^3}\,\mathbfit{r},
\end{equation}
in which $\xi$ is an arbitrary constant, the system~\eqref{Eq:two_body_problem} remains integrable. Perturbations of this type model various physical phenomena, like the solar radiation pressure acting on a surface perpendicular to the Sun vector, or simple control laws for spacecraft with solar electric propulsion. Indeed, the perturbed problem can be written
\begin{equation}
	\oder{^2\mathbfit{r}}{t^2} = - \frac{\mu(1-\xi)}{r^3}\mathbfit{r} = -\frac{\mu^\ast}{r^3}\,\mathbfit{r},
\end{equation}
which is equivalent to Kepler's problem with a modified gravitational parameter, $\mu^\ast = \mu(1-\xi)$.

The gravitational acceleration written in the intrinsic frame $\mathfrak{I}=\{\mathbfit{t},\mathbfit{n},\mathbfit{b}\}$, with
\begin{equation}
	\mathbfit{t} = \mathbfit{v}/v,\qquad \mathbfit{b}=\mathbfit{h}/h,\qquad \mathbfit{n} = \mathbfit{b}\times\mathbfit{t}
\end{equation} 
defined in terms of the velocity and angular momentum vectors, $\mathbfit{v}$ and $\mathbfit{h}$ respectively, reads
\begin{equation}\label{Eq:ag}
	\mathbfit{a}_g = -\frac{\mu}{r^3}\mathbfit{r} = -\frac{\mu}{r^2}(\cos\psi\,\mathbfit{t} - \sin\psi\,\mathbfit{n}).
\end{equation}
The flight-direction angle $\psi$ is given by
\begin{equation}
	\cos\psi = \frac{(\mathbfit{r}\cdot\mathbfit{v})}{rv}.
\end{equation}
Consequently, perturbations of the form $\mathbfit{a}_p=\xi\mu\,(\mathbfit{r}/r^3)$ result in
\begin{equation}\label{Eq:ap_general_1}
	\mathbfit{a}_p =  \frac{\xi\mu}{r^2}(\cos\psi\,\mathbfit{t} - \sin\psi\,\mathbfit{n}).
\end{equation}
Since the component normal to the velocity does not perform any work, \citet{bacon1959logarithmic} neglected its contribution and found that 
\begin{equation}
	\mathbfit{a}_p =  \frac{\mu}{2r^2}\cos\psi\,\mathbfit{t}\qquad 
\end{equation}
furnishes another integrable problem: the trajectory of the particle is a logarithmic spiral.

Motivated by this line of thought, we can treat the normal component of the acceleration in Eq.~\eqref{Eq:ap_general_1} separately by introducing a second parameter, $\eta$:
\begin{equation}\label{Eq:ap_original}
	\mathbfit{a}_p = \frac{\mu}{r^2}(\xi\cos\psi\,\mathbfit{t} + \eta\sin\psi\,\mathbfit{n}).
\end{equation}
The dimensionless parameters $\xi$ and $\eta$ are assumed constant. The forcing parameter $\xi$ controls the power exerted by the perturbation,
\begin{equation}
	\oder{\E_k}{t} = \mathbfit{a}_p\cdot\mathbfit{v} = \frac{\xi \mu}{r^2}\,v \cos\psi,
\end{equation}
with $\E_k$ the Keplerian energy of the system. The second parameter $\eta$ scales the terms that do not perform any work. Introducing the accelerations~\eqref{Eq:ag} and~\eqref{Eq:ap_original} into Eq.~\eqref{Eq:two_body_problem} yields
\begin{equation}\label{Eq:two_body_problem_projected}
	\oder{^2\mathbfit{r}}{t^2} = -\frac{\mu}{r^2}(1-\xi) (\cos\psi\,\mathbfit{t} - \gamma\sin\psi\,\mathbfit{n} ),
\end{equation}
having replaced $\eta$ by the new parameter
\begin{equation}
	\gamma = \frac{1+\eta}{1-\xi}.
\end{equation}

\begin{figure}
	\centering
	\includegraphics[width=5.5cm]{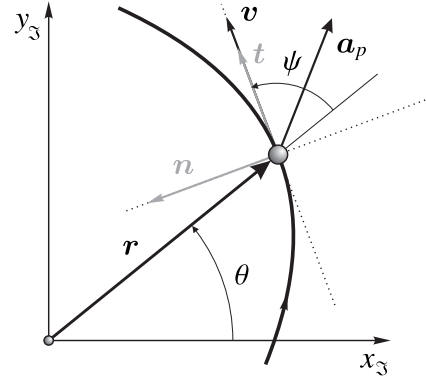}
	\caption{Geometry of the problem.\label{Fig:geometry}}
\end{figure}

Since the motion is planar we shall introduce a system of polar coordinates $(r,\theta)$ to formulate the problem. The radial velocity is defined as
\begin{equation}
	\dot{r} = \frac{(\mathbfit{v}\cdot\mathbfit{r})}{r} = v\cos\psi,
\end{equation}
whereas the projection of the velocity normal to the radiusvector takes the form
\begin{equation}
	r\dot{\theta} = v\sin\psi.
\end{equation}
Consequently, these geometric relations unveil the dynamical equations:
\begin{equation}
\begin{split}
	 \oder{r}{t} & = v\cos\psi \\
	 \oder{\theta}{t} & = \frac{v}{r}\sin\psi  \label{Eq:dthetadt_chi} \\
	 v & = \sqrt{\dot{r}^2+(r\dot{\theta})^2}.  
\end{split}
\end{equation}
The dynamics are referred to an inertial frame $\mathfrak{I}=\{\mathbfit{i}_\mathfrak{I},\mathbfit{j}_\mathfrak{I},\mathbfit{k}_\mathfrak{I}\}$, with $\mathbfit{k}_\mathfrak{I}\parallel\mathbfit{h}$ and using the $x_\mathfrak{I}$-axis as the origin of angles. Figure~\ref{Fig:geometry} depicts the configuration of the problem. We restrict the study to prograde motions ($\dot{\theta}>0$) without losing generality.

\subsection{Similarity transformation}
Consider a linear transformation $\mathscr{S}$ of the form
\begin{equation}
	\mathscr{S}:(t,r,\theta,\dot{r},\dot{\theta})\mapsto(\tau,\rho,\theta,\rho^\prime,\theta^\prime)
\end{equation}
defined explicitly by the positive constants $\alpha$, $\beta$, and $\delta=\alpha/\beta$:
\begin{equation}
	\tau = \frac{t}{\beta},\qquad  \rho = \frac{r}{\alpha}, \qquad \rho^\prime = \frac{\dot{r}}{\delta} , \qquad \theta^\prime = \beta\,\dot{\theta}.
\end{equation}
The constants $\alpha$, $\beta$, and $\delta$ have units of length, time, and velocity, respectively. The symbol $\square^\prime$ denotes derivatives with respect to $\tau$, whereas $\dot{\square}$ is reserved for derivatives with respect to $t$. The scaling factor $\alpha$ can be seen as the ratio of a homothetic transformation that simply dilates or contracts the orbit. Similarly, $\beta$ represents a time dilation or contraction. The velocity of the particle transforms into 
\begin{equation}
	\tilde{v} = \frac{v}{\delta}.
\end{equation}
In addition, $\beta$ and $\delta$ are defined in terms of $\alpha$ by virtue of
\begin{equation}
	\beta= \sqrt{\frac{\alpha^3}{ \mu\gamma(1-\xi)}} \qquad \text{and}\qquad \delta = \frac{\alpha}{\beta} = \sqrt{\frac{ \mu\gamma(1-\xi)}{\alpha}}.
\end{equation}
We assume $\gamma>0$ and $\xi<1$ for consistency. 

Equation~\eqref{Eq:two_body_problem_projected} then becomes
\begin{equation}\label{Eq:two_body_transformed1}
	\oder{^2\boldsymbol{\rho}}{\tau^2} = -\frac{1}{\rho^2}  \left(\frac{1}{\gamma}\cos\psi\,\mathbfit{t} - \sin\psi\,\mathbfit{n} \right),
\end{equation}
which is equivalent to the normalized two-body problem
\begin{equation}\label{Eq:two_body_transformed}
	\oder{^2\boldsymbol{\rho}}{\tau^2} = - \frac{\boldsymbol{\rho}}{\rho^3} + \tilde{\mathbfit{a}}_p
\end{equation}
perturbed by the purely tangential acceleration
\begin{equation}\label{Eq:ap_transformed}
	\tilde{\mathbfit{a}}_p = \frac{\gamma-1}{\gamma\rho^2}\cos\psi\,\mathbfit{t}.
\end{equation}
This result shows that $\mathscr{S}$ establishes a \emph{similarity transformation} between the two-body problem~\eqref{Eq:two_body_problem} perturbed by the acceleration~\eqref{Eq:ap_original}, and the simpler problem in Eq.~\eqref{Eq:two_body_transformed} and perturbed by~\eqref{Eq:ap_transformed}. That is, $\mathscr{S}^{-1}$ transforms the solution to Eq.~\eqref{Eq:two_body_transformed1}, $\boldsymbol{\rho}(\tau)$, into the solution to Eq.~\eqref{Eq:two_body_problem_projected}, $\mathbfit{r}(t)$.

Using $\mathbfit{q}=(\rho,\theta)$ and ${\mathbfit{q}}^\prime=(\rho^\prime,\theta^\prime)$ as the generalized coordinates and velocities, respectively, the dynamics of the problem abide by the Euler-Lagrange equations
\begin{equation}
	\oder{}{\tau}\left( \pder{\mathscr{L}}{q^\prime_i} \right) - \pder{\mathscr{L}}{q_i} = Q_i.
\end{equation} 
The Lagrangian of the transformed system takes the form
\begin{equation}\label{Eq:Lagrangian}
	\mathscr{L} = \frac{1}{2}({\rho^\prime}^2 + \rho^2{\theta^\prime}^2) + \frac{1}{\rho}
\end{equation}
and the generalized forces $Q_i$ read
\begin{equation}
	Q_\rho  = \frac{(\gamma-1)}{\gamma}\left( \frac{\rho^\prime}{\rho\tilde{v}} \right)^2 \quad \text{and}\quad  Q_\theta  = \frac{(\gamma-1)}{\gamma}  \left( \frac{ \rho^\prime\theta^\prime}{ \tilde{v}^2} \right).
\end{equation}

\subsection{Integrals of motion and dynamical symmetries} 
Let us introduce an infinitesimal transformation $\mathscr{R}$:
\begin{equation}\label{Eq:transformation_infinitesimal}
\begin{split}
	 \tau &\to \tau^\ast = \tau + \varepsilon f(\tau;q_i,q^\prime_i) \\
	 q_i & \to q_i^\ast = q_i + \varepsilon F_i(\tau;q_i,q^\prime_i),
\end{split}
\end{equation}
defined in terms of a small parameter $\varepsilon\ll1$ and the generators $F_i$ and $f$. For transformations that leave the action unchanged up to an exact differential,
\begin{equation}
	\mathscr{L}\left(\tau^\ast;q_i^\ast,\oder{q_i^\ast}{\tau^\ast}\right)\mathrm{d}\tau^\ast - \mathscr{L}\left(\tau ;q_i ,\oder{q_i }{\tau}\right)\mathrm{d}\tau = \varepsilon\,\mathrm{d}\Psi(\tau;q_i,q_i^\prime),
\end{equation}
with $\Psi(\tau;q_i,q_i^\prime)$ a given gauge, Noether's theorem  states that
\begin{equation}\label{Eq:Noether}
	\sum_i\left\{ \left(\pder{\mathscr{L}}{q^\prime_i}\right)F_i + f\left[ \mathscr{L} - \left(\pder{\mathscr{L}}{q^\prime_i}\right)q^\prime_i \right] \right\} - \Psi(\tau;q_i,q^\prime_i) = \Lambda
\end{equation}
is a first integral of the problem. Here $\Lambda$ is a certain constant of motion. We refer the reader to the work of \citet{efroimsky2004gauge} for a refreshing look into the role of gauge functions in celestial mechanics.

Since the perturbation in Eq.~\eqref{Eq:ap_transformed} is not conservative, we shall focus on the extension of Noether's theorem to nonconservative systems by \citet{djukic1975noether}. It must be $F_i-q^\prime_if\neq0$ for the conservation law to hold \citep{vujanovic1986some}. For the case of nonconservative systems the generators $F_i$, $f$, and the gauge $\Psi$ need to satisfy the following relation:
\begin{align}
	   \sum_i\Bigg\{ \left( \pder{\mathscr{L}}{q_i} \right) F_i  + \left(\pder{\mathscr{L}}{q^\prime_i} \right)( {F}^\prime_i & - q^\prime_i{f}^\prime) + Q_i(F_i-q^\prime_if)\Bigg\} \nonumber\\
	&   + f^\prime \mathscr{L} + f\pder{\mathscr{L}}{\tau} = {\Psi}^\prime. \label{Eq:Noether_identity}
\end{align}
This equation and the condition $F_i-q^\prime_if\neq0$ furnish the generalized Noether-Bessel-Hagen (NBH) equations \citep{trautman1967noether,djukic1975noether,vujanovic1986some}. The NBH equations involve the full derivative of the gauge function and the generators with respect to $\tau$, meaning that Eq.~\eqref{Eq:Noether_identity} depends on the partial derivatives of $\Psi$, $F_i$, and $f$ with respect to time, the coordinates, and the velocities. By expanding the convective terms the NBH equations decompose in the system of Killing equations:
\begin{alignat}{2}
	& \mathscr{L} \pder{f}{q_j^\prime} +\sum_i \pder{\mathscr{L}}{q^\prime_i}\left( \pder{F_i}{q_j^\prime} - q^\prime_i\pder{f}{q_j^\prime } \right)  = \pder{\Psi}{q_j^\prime}, \nonumber\\[2mm]
	& \pder{}{\tau}(f\mathscr{L} - \Psi) + \sum_i \Bigg\{   \pder{\mathscr{L}}{q_i}F_i + \mathscr{L}\pder{f}{q_i}q_i^\prime + Q_i(F_i-q_i^\prime f) \nonumber\\
	& + \pder{\mathscr{L}}{q_i^\prime}\left[ \pder{F_i}{\tau} - q_i^\prime\pder{f}{\tau} + \sum_j\left( \pder{F_i}{q_j}q_j^\prime - q_i^\prime q_j^\prime\pder{f}{q_j} \right) \right] - \pder{\Psi}{q_i}q_i^\prime   \Bigg\} =  0. \label{Eq:killing2} 
\end{alignat}
The system~\eqref{Eq:killing2} decomposes in three equations that can be solved for the generators $F_\rho$, $F_\theta$, and $f$ given a certain gauge. If the transformation defined in Eq.~\eqref{Eq:transformation_infinitesimal} satisfies the NBH equations, then the system admits the integral of motion~\eqref{Eq:Noether}.

\subsubsection{Generalized equation of the energy}
The Lagrangian in Eq.~\eqref{Eq:Lagrangian} is time-independent. Thus, the action is not affected by arbitrary time transformations. In the Keplerian case a simple time translation reveals the conservation of the energy. Motivated by this fact, we explore the generators
\begin{equation}
	f = 1, \qquad F_\rho=0, \qquad \text{and} \qquad F_\theta=0.
\end{equation}
Solving Killing equations~\eqref{Eq:killing2} with the above generators leads to the gauge function
\begin{equation}
	\Psi = \frac{\gamma-1}{\gamma r}.	
\end{equation}
Provided that the NBH equations hold, the system admits the integral of motion
\begin{equation}\label{Eq:integr1}
	\frac{\tilde{v}^2}{2} - \frac{1}{\gamma {\rho}}  = -\Lambda \equiv \frac{\tka}{2},
\end{equation}
written in terms of the constant $\tka=-2\Lambda$. This term can be solved from the initial conditions
\begin{equation}
	\tka = \tilde{v}_0^2 - \frac{2}{\gamma {\rho}_0}.
\end{equation}
When $\gamma=1$ the perturbation~\eqref{Eq:ap_transformed} vanishes and Eq.~\eqref{Eq:integr1} reduces to the normalized equation of the Keplerian energy. In fact, in this case $\tka$ becomes twice the Keplerian energy of the system, $\tka=2\tilde{\E}_k$. Moreover, the gauge vanishes and Eq.~\eqref{Eq:Noether} furnishes the Hamiltonian of Kepler's problem. The integral of motion~\eqref{Eq:integr1} is a generalization of the equation of the energy. 

In the Keplerian case ($\gamma=1$ and $\xi=0$) the sign of the energy determines the type of solution. Negative values of $\tka$ yield elliptic orbits, positive values correspond to hyperbolas, and the orbits are parabolic for vanishing $\tka$. We shall extend this classification to the general case $\gamma\neq 1$: the solutions will be classified as elliptic ($\tka<0$), parabolic ($\tka=0$), and hyperbolic ($\tka>0$) orbits.

\subsubsection{Generalized equation of the angular momentum}

In the unperturbed problem $\theta$ is an ignorable coordinate. Indeed, a simple translation in $\theta$ (a rotation) with $f=F_\rho=\Psi=0$ and $F_\theta=1$ yields the conservation of the angular momentum. In order to extend this first integral to the perturbed case, we consider the same generator $F_\theta=1$. However, solving for the gauge and the remaining generators in Killing equations yields the nontrivial functions
\begin{equation}
\begin{split}
	 F_\rho & = \frac{\rho^\prime}{\theta^\prime}(1-\tilde{v}^{\gamma-1}), \\[1mm]
	 f   & = \frac{1-v^{\gamma-1}}{\theta^\prime} + (1-\gamma)\rho^2\theta^\prime\tilde{v}^{\gamma-3},\\
	 \Psi & = \frac{1}{2\rho\theta^\prime}\Big\{ \tilde{v}^2\left[ \rho - (3-\gamma)\rho\tilde{v}^{\gamma-1} \right] + 2 - \tilde{v}^{\gamma-1}[2\gamma - (3-\gamma){\rho^\prime}^2\rho] \\
	  &\qquad\qquad + \tilde{v}^{\gamma-3}[ \rho(\rho^4{\theta^\prime}^4-{\rho^\prime}^4) - 2(1-\gamma){\rho^\prime}^2 ] \Big\}.
\end{split}
\end{equation}
They satisfy $F_i-q_i^\prime\neq0$. Noether's theorem holds and Eq.~\eqref{Eq:Noether} furnishes the integral of motion
\begin{equation}\label{Eq:integr2}
	\rho^2\tilde{v}^{\gamma-1}\theta^\prime = \Lambda \equiv \tkb.
\end{equation}
This first integral is none other than a generalized form of the conservation of the angular momentum. Indeed, making $\gamma=1$ Eq.~\eqref{Eq:integr2} reduces to
\begin{equation}
	\rho^2\theta^\prime = \tkb,
\end{equation}
where $\tkb$ coincides with the angular momentum of the particle. In addition, the generators $F_\rho$ and $f$, and the gauge vanish when $\gamma$ equals unity. 

By recovering Eq.~\eqref{Eq:dthetadt_chi} the integral of motion~\eqref{Eq:integr2} can be written:
\begin{equation}
	\rho\tilde{v}^\gamma\sin\psi = \tkb
\end{equation}
in terms of the coordinates intrinsic to the trajectory. The fact that $\sin\psi\leq1$ forces 
\begin{equation}
	\tkb\leq\tilde{v}^\gamma{\rho}\implies \tkb^2\leq\tilde{v}^{2\gamma}{\rho}^2.
\end{equation}
The second step is possible because all variables are positive. Combining this expression with Eq.~\eqref{Eq:integr1} yields
\begin{equation}\label{Eq:ineq}
	\tkb^2\leq\left(\tka + \frac{2}{\gamma{\rho}}\right)^\gamma{\rho}^2.
\end{equation}
Depending on the values of $\gamma$, Eq.~\eqref{Eq:ineq} may define upper or lower limits to the values that the radius ${\rho}$ can reach. In general, this condition can be resorted to provide the polynomial constraint
\begin{equation}
	P_\mathrm{nat}(\rho) \geq 0,
\end{equation} 
where $P_\mathrm{nat}(\rho)$ is a polynomial of degree $\gamma$ in $\rho$ whose roots dictate the nature of the solutions. This inequality will be useful for defining the different families of orbits.

\subsection{Properties of the similarity transformation}
The main property of the similarity transformation $\mathscr{S}$ is that it does not change the type of the solution, i.e. the sign of $\tka$ is not altered. Applying the inverse similarity transformation $\mathscr{S}^{-1}$ to Eq.~\eqref{Eq:integr1} yields
\begin{equation}
	\tka = \frac{v^2}{\delta^2} - \frac{2\alpha}{\gamma r} = \frac{v^2\alpha}{\mu\gamma(1-\xi)} - \frac{2\alpha}{\gamma r} = \frac{\alpha}{\mu\gamma(1-\xi)}\left[ v^2 - \frac{2\mu}{r}(1-\xi) \right]  
\end{equation}
and results in
\begin{equation}
	\ka = \delta^2\,\tka =  v^2  - \frac{2\mu}{r}(1-\xi) = v^2 - \frac{2\mu^\ast}{r}.
\end{equation}
Since $\delta^2>0$ no matter the values of $\xi$ or $\gamma$, the sign of $\ka$ is not affected by the transformation $\mathscr{S}$. If the solution to the original problem~\eqref{Eq:two_body_problem} is elliptic, the solution to the reduced problem~\eqref{Eq:two_body_transformed} will be elliptic too, and vice-versa. The transformation reduces to a series of scaling factors affecting each variable independently. 

The integral of the angular momentum transforms into
\begin{equation}
	\tkb = \frac{r^2v^{\gamma-1}\dot{\theta}}{\alpha\delta^{\gamma}} = \frac{\kb}{\alpha\delta^\gamma} \implies \kb = \alpha\delta^\gamma\tkb.
\end{equation}
The constant $\tkb$ remains positive, although the scaling factor $\alpha\delta^\gamma$ can modify its value significantly. 

The transformation is defined in terms of three parameters: $\alpha$, $\xi$ and $\gamma$. For $\xi=\xi_\gamma$, with
\begin{equation}
	\xi_\gamma = 1 - \frac{\alpha}{\gamma},
\end{equation}
$\mathscr{S}$ reduces to the identity map. Choosing $\alpha=1$ the special values of $\xi$ that yield trivial transformations for $\gamma=1$, $2$, $3$ and $4$ are, respectively, $\xi_\gamma=0$, $1/2$, $2/3$ and $3/4$. The similarity transformation can be understood from a different approach: solving the simplified problem~\eqref{Eq:two_body_transformed1} is equivalent to solving the full problem~\eqref{Eq:two_body_problem_projected}, but setting $\xi=\xi_\gamma$.

%\subsection{Families of solutions}

Combining Eqs.~\eqref{Eq:dthetadt_chi} renders
\begin{equation}\label{Eq:to_integrate}
	\oder{\theta}{{\rho}} = \frac{\tan\psi}{{\rho}}.
\end{equation}
The right-hand side of this equation can be written as a function of ${\rho}$ alone thanks to 
\begin{equation}
	\tan\psi = \frac{n\tkb}{\sqrt{(\tilde{v}^\gamma{\rho})^2-\tkb^2}}.
\end{equation}
The parameter $n$ is $n=+1$ for orbits in a raising regime ($\dot{r}>0$), and $n=-1$ for a lowering regime ($\dot{r}<0$). The velocity is solved from Eq.~\eqref{Eq:integr1}. Integrating Eq.~\eqref{Eq:to_integrate} furnishes the solution $\theta({\rho})$, which can then be inverted to define the trajectory ${\rho}(\theta)$.

\subsection{Solvability}
The trajectory of the particle is obtained upon integration and inversion of Eq.~\eqref{Eq:to_integrate}. This equation can be written
\begin{equation}\label{Eq:dthetadr_integrability}
	\oder{\theta}{{\rho}} = \frac{n\tkb}{\sqrt{P_\mathrm{sol}({\rho})}},
\end{equation}
where $P_\mathrm{sol}({\rho})$ is a polynomial in ${\rho}$, in particular
\begin{equation}
	P_\mathrm{sol}(\rho) = \rho^2 P_\mathrm{nat}(\rho).
\end{equation}
The roots of $P_\mathrm{sol}(\rho)$ determine the \emph{form} of the solution and coincide with those of $P_\mathrm{nat}(\rho)$ (obviating the trivial ones). The integration of Eq.~\eqref{Eq:dthetadr_integrability} depends on the factorization of $P_\mathrm{sol}(\rho)$. This polynomial expression can be expanded thanks to the binomial theorem:
\begin{equation}
	P_\mathrm{sol}({\rho})=\sum_{k=0}^\gamma\left(\!\begin{array}{c}
		\gamma\\[1mm]
		k
	\end{array}\!\right) \frac{2^k}{\gamma^k}{\rho}^{4-k}\tka^{\gamma-k} - {\rho}^2\tkb
\end{equation}
with $\gamma\neq0$ an integer. For $\gamma\leq4$ the polynomial is of degree four: when $\gamma=1$ or $\gamma=2$ there are two trivial roots and Eq.~\eqref{Eq:dthetadr_integrability} can be integrated using elementary functions; when $\gamma=3$ or $\gamma=4$ it yields elliptic integrals. Negative values of $\gamma$ or positive values greater than four lead to a polynomial $P_\mathrm{sol}({\rho})$ with degree five or above. The solution can no longer be reduced to elementary functions nor elliptic integrals \citep[p.~512]{whittaker1927course}, for it is given by hyperelliptic integrals. This special class of Abelian integrals can only be inverted in very specific situations \citep[see][pp.~252--271]{byrd1954handbook}. Thus, we shall focus on the solutions to the cases $\gamma=1$, $2$, $3$ and $4$. 

The following sections~\ref{Sec:conic}--\ref{Sec:sinusoidal} present the corresponding families of orbits. For $\gamma=1$ the solutions to the reduced problem are Keplerian orbits. For the case $\gamma=2$ the solutions are called \emph{generalized logarithmic spirals}, because for $\tka=0$ the particle describes a logarithmic spiral \citep{roa2016new}. The cases $\gamma=3$ and $\gamma=4$ yield \emph{generalized cardioids} and \emph{generalized sinusoidal spirals}, respectively (for $\tka=0$ the orbits are cardioids and sinusoidal spirals).

%After characterizing the orbits, Sec.~\ref{Sec:Weierstrass} presents a unified form of the solutions to the cases $\gamma=3$ and $\gamma=4$ introducing the Weierstrass elliptic functions. 

\section{Case $\gamma=1$: conic sections}\label{Sec:conic}
For $\gamma=1$ the integrals of motion~\eqref{Eq:integr1} and \eqref{Eq:integr2} reduce to the normalized equations of the energy and angular momentum, respectively. The condition on the radius given by Eq.~\eqref{Eq:ineq} becomes
\begin{equation}\label{Eq:polyn_g1}
	P_\mathrm{nat}(\rho)\equiv \tka{\rho}^2 + {2}{\rho} -\tkb^2  \geq0.
\end{equation}
For the case $\tka<0$ (elliptic solution) this translates into ${\rho}\in[{\rho}_\mathrm{min},{\rho}_\mathrm{max}]$, where
\begin{equation}
\begin{split}
	{\rho}_\mathrm{min} & = \frac{1}{(-\tka)}\left(1-\sqrt{1+\tka\tkb^2}\right)  \\
	{\rho}_\mathrm{max} & = \frac{1}{(-\tka)}\left(1+\sqrt{1+\tka\tkb^2}\right)  .
\end{split}
\end{equation}
These limits are none other than the periapsis and apoapsis radii, provided that $\tka$ and $\tkb$ relate to the semimajor axis and eccentricity by means of:
\begin{equation}
	\frac{1}{(-\tka)} = \tilde{a}\qquad \text{and}\qquad \sqrt{1+\tka\tkb^2} = \tilde{e}.
\end{equation}
For $\tka=0$ (parabolic case) the semimajor axis becomes infinite, and Eq.~\eqref{Eq:polyn_g1} has only one root corresponding to ${\rho}=\tkb^2/2$. Note that it must be $\tkb^2>-1/\tka$. Similarly, for hyperbolic orbits ($\tka>0$) it is ${\rho}\geq {\rho}_\mathrm{min}$ as ${\rho}_\mathrm{max}$ becomes negative.

Thus, the solution is simply a conic section:
\begin{equation}\label{Eq:traj_Kepler}
	{\rho}(\theta) = \frac{\tilde{h}^2}{1+\tilde{e}\cos(\theta-\theta_m)} = \frac{\tkb^2}{1+\sqrt{1+\tka\tkb^2}\,\cos(\theta-\theta_m)}.
\end{equation}
The angle $\theta_m=\varOmega+\omega$ defines the direction of the line of apses in frame $\mathfrak{I}$, meaning that $\theta-\theta_m$ is the true anomaly. If $\tka<0$, then ${\rho}(\theta_m)={\rho}_\mathrm{min}$, and ${\rho}(\theta_m+\piup)={\rho}_\mathrm{max}$. The velocity $\tilde{v}$ follows from the integral of the energy:
\begin{equation}
	\tilde{v} = \sqrt{\tka + {2}/{\rho}}.
\end{equation}
It is minimum at apoapsis and maximum at periapsis.

Applying the similarity transformation $\mathscr{S}^{-1}$ to the previous solution leads to the extended integral
\begin{equation}
	\frac{\ka}{2} = \delta^2\frac{\tka}{2}  = \frac{v^2}{2} - \frac{\mu}{r}(1-\xi)  = \frac{v^2}{2} - \frac{\mu^\ast}{r} .
\end{equation}
The factor $\mu^\ast=\mu(1-\xi)$ behaves as a modified gravitational parameter. This kind of solutions arise from, for example, the effect of the solar radiation pressure directed along the Sun-line on a particle following a heliocentric orbit \citep[][p.~121]{mcinnes2004solar}. 

%It is interesting to remark that the energy is conserved in the reduced problem, but not in the similar

\section{Case $\gamma=2$: generalized logarithmic spirals}
The case $\gamma=2$ yields the family of generalized logarithmic spirals found by \citet{roa2016new} in the context of interplanetary mission design. An extended version and the solution to the spiral Lambert problem can be found in following sequels \citep{roa2016introducing,roa2016spiral}. Negative values of $\tka$ define the so called elliptic spirals, positive values yield hyperbolic spirals, and the limit case $\tka=0$ corresponds to parabolic spirals, which turn out to be pure logarithmic spirals.
 
\begin{figure*} 
	\centering
	\subfloat[Elliptic\label{Fig:logspirals_elliptic}]{\includegraphics[width=.19\linewidth]{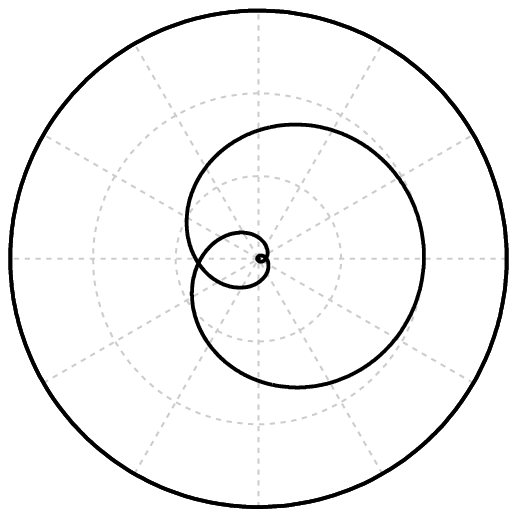}}\hspace{1mm}
	\subfloat[Parabolic\label{Fig:logspirals_parabolic}]{\includegraphics[width=.19\linewidth]{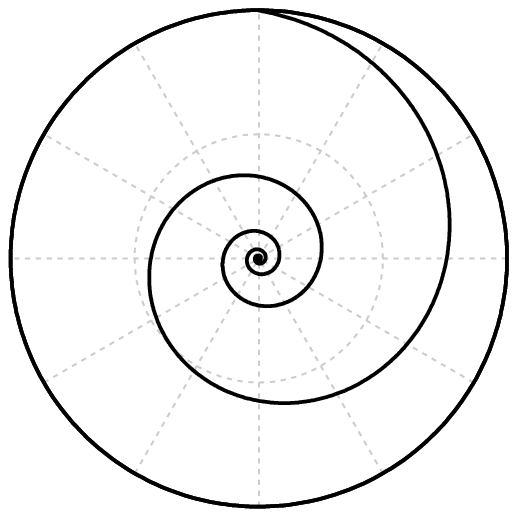}}\hspace{1mm}
	\subfloat[Hyperbolic Type I\label{Fig:logspirals_hypI}]{\includegraphics[width=.19\linewidth]{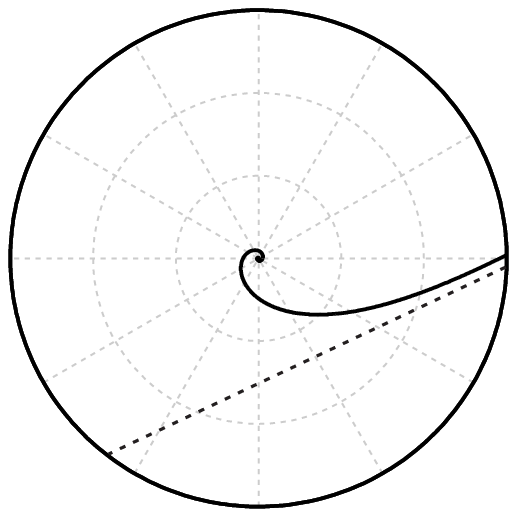}}\hspace{1mm}
	\subfloat[Hyperbolic Type II\label{Fig:logspirals_hypII}]{\includegraphics[width=.19\linewidth]{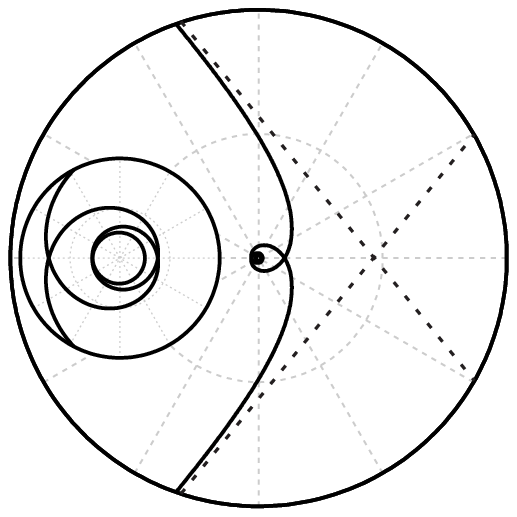}}\hspace{1mm}
	\subfloat[Hyperbolic transition\label{Fig:logspirals_hyptrans}]{\includegraphics[width=.19\linewidth]{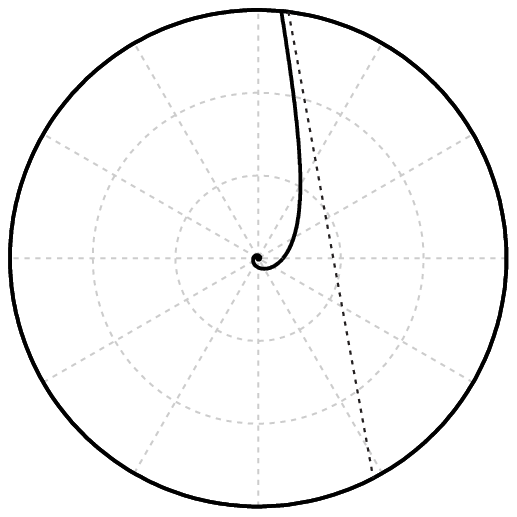}}
	\caption{Examples of generalized logarithmic spirals $(\gamma=2)$. A zoomed view of the periapsis of Type II hyperbolic spirals is included. \label{Fig:gen_log_spirals}}
\end{figure*}

\subsection{Elliptic motion}
For the case $\tka<0$ the inequality in Eq.~\eqref{Eq:ineq} translates into ${\rho}<{\rho}_\mathrm{max}$, with
\begin{equation}
	{\rho}_\mathrm{max} = \frac{1-\tkb}{(-\tka)}.
\end{equation}
Here ${\rho}_\mathrm{max}$ behaves as the apoapsis of the elliptic spiral. In addition, it forces $\tkb\leq1$. Spirals of this type initially in raising regime will reach ${\rho}_\mathrm{max}$, then transition to lowering regime and fall toward the origin. If they are initially in lowering regime the radius will decrease monotonically. The velocity at apoapsis is the minimum possible velocity, and reads
\begin{equation}
	\tilde{v}_m = \sqrt{\frac{\tkb}{{\rho}_\mathrm{max}}} = \sqrt{\frac{-\tka\tkb}{1-\tkb}}.
\end{equation}

Introducing the \emph{spiral anomaly} $\nu$:
\begin{equation}\label{Eq:spiral_anomaly}
	\nu(\theta) = \frac{\ell}{\tkb}(\theta-\theta_m)
\end{equation}
and with $\ell=\sqrt{1-\tkb^2}$, Eq.~\eqref{Eq:to_integrate} can be integrated and inverted to provide the equation of the trajectory:
\begin{equation}\label{Eq:traj_ellip_gamma2}
	\frac{{\rho}(\theta)}{{\rho}_\mathrm{max}} = \frac{1+\tkb}{1+\tkb\cosh\nu }.
\end{equation}
The angle $\theta_m$ defines the orientation of the apoapsis, i.e. ${\rho}(\theta_m)={\rho}_\mathrm{max}$. It can be solved form the initial conditions:
\begin{equation}
	\theta_m = \theta_0 + \frac{n\tkb}{\ell}\left|\arccosh\left[ -\frac{1}{\tkb}\left(1+\frac{\ell^2}{\tka{\rho}_0}\right) \right]\right|.
\end{equation}
The trajectory is symmetric with respect to $\theta_m$, provided that ${\rho}(\theta_m+\Delta\theta)={\rho}(\theta_m-\Delta\theta)$, and it is plotted in Fig.~\ref{Fig:logspirals_elliptic} \citep[see][Chap.~9, for details on the symmetry properties]{roa2016regularization}.
 
\subsection{Parabolic motion: the logarithmic spiral}
For parabolic spirals Eq.~\eqref{Eq:ineq} reduces to $\tkb\leq 1$, meaning that there are no limit radii. Spirals in raising regime escape to infinity, and spirals in lowering regime fall to the origin. The limit $\lim_{{\rho}\to\infty}\tilde{v} = 0$ shows that the particle reaches infinity along a spiral branch.

The particle follows a pure logarithmic spiral,
\begin{equation}\label{Eq:traj_parab_gamma2}
	{\rho}(\theta) = {\rho}_0\,\mathrm{e}^{(\theta-\theta_0)\cot\psi},
\end{equation}
keeping in mind that $\cot\psi=n\ell/\tkb$. See Fig.~\ref{Fig:logspirals_parabolic} for an example. In the reduced problem the velocity matches the local circular velocity, $\tilde{v} = \sqrt{1/\rho}$, and the parameter $\xi$ modifies the velocity in the full problem:
\begin{equation}
	v  = \delta \tilde{v} = \sqrt{\frac{2\mu(1-\xi)}{ r}} = \sqrt{\frac{2\mu^\ast}{ r}}.
\end{equation}
Note that when $\xi=1/2$ (or $\mu^\ast=\mu/2$) this is the true circular velocity.

\subsection{Hyperbolic motion}
Under the assumption $\tka>0$ the polynomial constraint in Eq.~\eqref{Eq:ineq} transforms into ${\rho}\geq{\rho}_\mathrm{min}$, with
\begin{equation}
	{\rho}_\mathrm{min} = \frac{\tkb-1}{\tka}.
\end{equation}
This equation yields two different cases: when $\tkb\leq1$ the periapsis radius ${\rho}_\mathrm{min}$ becomes negative, which means that the spiral reaches the origin when in lowering regime. Conversely, for $\tkb>1$ there is an actual periapsis; a spiral in lowering regime will reach ${\rho}_\mathrm{min}$, then transition to raising regime and escape. Hyperbolic spirals with $\tkb<1$ are of Type I, whereas $\tkb>1$ defines hyperbolic spirals of Type II.

Hyperbolic spirals reach infinity with a finite, nonzero velocity $\lim_{{\rho}\to\infty}\tilde{v} = \tilde{v}_\infty \equiv \sqrt{\tka}$. That is, the generalized constant of the energy $\tka$ is equivalent to the characteristic energy $\tilde{v}_\infty^2={C}_3$.

\subsubsection{Hyperbolic spirals of Type I}
The trajectory described by a hyperbolic spiral with $\tkb<1$ (Type I) takes the form
\begin{equation}\label{Eq:traj_hyperI_gamma2}
	{\rho}(\theta) = \frac{\ell^2/\tka}{2\sinh\frac{\nu}{2}\left( \sinh\frac{\nu}{2} + \ell\cosh\frac{\nu}{2} \right)}.
\end{equation}
In this case the spiral anomaly $\nu(\theta)$ reads
\begin{equation}
	\nu(\theta) = \frac{n\ell}{\tkb}(\theta_{\mathrm{as}}-\theta),
\end{equation}
where the direction of the asymptote is solved from
\begin{equation}
	\theta_\mathrm{as} = \theta_0 + \frac{n\tkb}{\ell}\ln\left[ \frac{\tkb(\ell|\cos\psi_0|+1-\tkb\sin\psi_0)}{(\tkb-\sin\psi_0)(1+\ell)} \right].
\end{equation}
An example of a Type I hyperbolic spiral connecting the origin with infinity can be found in Fig.~\ref{Fig:logspirals_hypI}. The dashed line represents the asymptote.

\subsubsection{Hyperbolic spirals of Type II}
Hyperbolic spirals of Type II are defined by
\begin{equation}\label{Eq:traj_hypII}
	\frac{{\rho}(\theta)}{{\rho}_\mathrm{min}} = \frac{1+\tkb}{1+\tkb\cos\nu }.
\end{equation}
The spiral anomaly $\nu(\theta)$ is defined in Eq.~\eqref{Eq:spiral_anomaly}, with $\ell^2=\tkb^2-1$. The line of apses follows the direction of
\begin{equation}
	\theta_m = \theta_0 - \frac{n\tkb}{\ell}\left| \arccos\left[ \frac{1}{\tkb}\left( \frac{\ell^2}{\tka{\rho}_0}-1 \right) \right] \right|.
\end{equation}
Equation~\eqref{Eq:traj_hypII} depends on the spiral anomaly by means of $\cos\nu$. Thus, the trajectory is symmetric with respect to $\theta_m$: the apse line is the axis of symmetry. The symmetry of the trajectory proves that there are two values of $\nu$ that cancel the denominator: there are two asymptotes, defined explicitly by
\begin{equation}
	\theta_\mathrm{as} = \theta_m \pm \frac{\tkb}{\ell}\arccos\left(-\frac{1}{\tkb}\right).%,\quad  \theta_\mathrm{as,2} = \theta_m - \frac{\tkb}{\ell}\arccos\left(-\frac{1}{\tkb}\right).
\end{equation}
Both asymptotes are symmetric with respect to the line of apses. The shape of the hyperbolic spirals of Type II can be analyzed in Fig.~\ref{Fig:logspirals_hypII}. The figure includes a zoomed view of the periapsis region.

% The figure includes three curves, corresponding to elliptic, parabolic, and Type II hyperbolic spirals. Elliptic spirals are bounded, symmetric, and after reaching the apoapsis radius they transition to lowering regime and fall to the origin. Parabolic spirals are simply logarithmic spirals. Hyperbolic spirals of Type II exhibit two symmetric asymptotes, with the line of apses aligned with the periapsis.Hyperbolic spirals of Type I are not depicted in the figure because they simply connect the origin with infinity. 

\subsubsection{Transition between Type I and Type II hyperbolic spirals}
Hyperbolic spirals of Type I have been defined for $\tkb<1$, whereas $\tkb>1$ yields hyperbolic spirals of Type II. In the limit case $\tkb=1$ the equations of motion simplify noticeably; the resulting spiral is
\begin{equation}\label{Eq:traj_hyperlim_gamma2}
	{\rho}(\theta) = \frac{2/\tka}{\nu (\nu+2n)}.
\end{equation}
The angular variable $\nu(\theta)$ is defined with respect to the orientation of the asymptote, i.e.
\begin{equation}
	\nu(\theta) = \theta_\mathrm{as} - \theta.
\end{equation}
The asymptote is fixed by
\begin{equation}
	\theta_\mathrm{as} = \theta_0 - n\left( 1-\sqrt{1+\frac{2}{\tka {\rho}_0}} \right).
\end{equation}
Qualitatively, this type of solution is equivalent to a Type II spiral with $\rho_\mathrm{min}\to0$. The trajectory is similar to Type I spirals, or to one half of a Type II spiral (see Fig.~\ref{Fig:logspirals_hyptrans}).

\section{Case $\gamma=3$: Generalized cardioids}
The condition in Eq.~\eqref{Eq:ineq}, yields the polynomial inequality:
\begin{equation}\label{Eq:polynomial_gamma3}
	P_\mathrm{nat}(\rho)\equiv\tka^3\,{\rho}^3 + 2\tka^2\,{\rho}^2 + \left(\frac{4\tka}{3}-\tkb^2\right){\rho} + \frac{8}{27} \geq 0 .
\end{equation}
The discriminant $\Delta$ of the polynomial $P_\mathrm{nat}(\rho)$ predicts the nature of the roots. It is 
\begin{equation}\label{Eq:discriminant}
	\Delta = -4\tka^3\tkb^4(3\tka-\tkb^2).
\end{equation}
The intermediate value theorem shows that there is at least one real root. For the elliptic case $(\tka<0)$ it is $\Delta<0$, meaning that the other two roots are complex conjugates. In the hyperbolic case $(\tka>0)$ the sign of the discriminant depends on the values of $\tkb$: if $\tkb^2>3\tka$ it is $\Delta>0$, and for $\tkb^2<3\tka$ the discriminant is negative. This behavior yields two types of hyperbolic solutions.

\subsection{Elliptic motion}
The nature of elliptic motion is determined by the polynomial constraint in Eq.~\eqref{Eq:polynomial_gamma3}. The only real root is given by
\begin{equation}\label{Eq:rapo_gamma3}
	{\rho}_1 = \frac{\Lambda(\Lambda+2\tka^2)+3\tka^3\tkb^2}{3(-\tka)^3\Lambda}
	%
	%{\rho}_1 = \frac{\Lambda\tka^2(\Lambda\tka+2)+3\tkb^2}{3(-\tka)^3\Lambda}
\end{equation}
with
\begin{equation}\label{Eq:Lambda}
	\Lambda = (-\tka)\left\{ 3(-\tka)\tkb^2\left[ \sqrt{-3\tka(\tkb^2-3\tka)} + 3\tka \right] \right\}^{1/3}.
\end{equation}
Equation~\eqref{Eq:polynomial_gamma3} reduces to ${\rho}-{\rho}_1\leq0$ and we shall write 
\begin{equation}
	{\rho}_\mathrm{max}\equiv{\rho}_1.
\end{equation}
Thus, elliptic generalized cardioids never escape to infinity because they are bounded by $\rho_\mathrm{max}$. When in raising regime they reach the apoapsis radius ${\rho}_\mathrm{max}$, then transition to lowering regime and fall toward the origin. The velocity at apoapsis,
\begin{equation}
	\tilde{v}_m = \sqrt{\tka + {2}/({3{\rho}_\mathrm{max}})},
\end{equation}
is the minimum velocity in the cardioid.

Equation~\eqref{Eq:to_integrate} can be integrated from the initial radius $r_0$ to the apoapsis of the cardioid, and the result provides the orientation of the line of apses:
%\begin{equation}
%	\theta({\rho}) - \theta_m = \frac{\tkb\EllF(\phi,k)}{(-\tka)^{3/2}g}
%\end{equation}
%
\begin{equation}
	\theta_m = \theta_0 +  \frac{n\tkb}{\sqrt{AB}(-\tka)^{3/2}}[2\EllK(k) - \EllF(\phi_0,k)],
\end{equation}
where $\EllK(k)$ and $\EllF(\phi_0,k)$ are the complete and incomplete elliptic integrals of the first kind, respectively. Introducing the auxiliary term
\begin{equation}
	\lambda_{ij} = \rho_i - \rho_j
\end{equation}
their argument and modulus read
\begin{equation}
	\phi_0 = \arccos\left[  \frac{B\lambda_{10}-A{\rho}_0}{B\lambda_{10} + A{\rho}_0}  \right],\quad k = \sqrt{\frac{{\rho}_\mathrm{max}^2 - (A-B)^2}{4AB}}.
\end{equation}
The previous definitions involve the auxiliary parameters:
\begin{equation}
	A = \sqrt{({\rho}_\mathrm{max}-b_1)^2+a_1^2},\qquad B = \sqrt{b_1^2+a_1^2}
\end{equation}
and 
\begin{equation}
	%b_1 = \frac{\Lambda\tka^2(\Lambda\tka-4)+3\tkb^2}{6\Lambda\tka^3},\quad a_1^2 = \frac{(\Lambda^2\tka^3-3\tkb^2)^2}{12\Lambda^2\tka^6} 
	%
	b_1 = \frac{\Lambda(\Lambda-4\tka^2)+3\tka^3\tkb^2}{6\tka^3\Lambda},\qquad a_1^2 = \frac{(\Lambda^2-3\tka^3\tkb^2)^2}{12\tka^6\Lambda^2}.
\end{equation}
Recall the definition of $\Lambda$ in Eq.~\eqref{Eq:Lambda}.

The equation of the trajectory is obtained by inverting the function $\theta({\rho})$, and results in:
\begin{equation}\label{Eq:ellip_g3}
	\frac{{\rho}(\theta)}{{\rho}_\mathrm{max}} = \left\{ 1 + \frac{A}{B}\left[ \frac{1-\cn(\nu ,k)}{1+\cn(\nu ,k)}  \right] \right\}^{-1}.
\end{equation}
It is defined in terms of the Jacobi elliptic function $\cn(\nu,k)$. The anomaly $\nu$ reads
\begin{equation}
	\nu(\theta) = \frac{(-\tka)^{3/2}}{\tkb}\sqrt{AB}\,(\theta-\theta_m).
\end{equation}
Equation~\eqref{Eq:ellip_g3} is symmetric with respect to the apse line, ${\rho}(\theta_m+\Delta\theta)={\rho}(\theta_m-\Delta\theta)$, as shown in Fig.~\ref{Fig:cardioids_ellip}. 

\subsection{Parabolic motion: the cardioid}
When the constant of the generalized energy $\tka$ vanishes the condition in Eq.~\eqref{Eq:ineq} translates into ${\rho}<{\rho}_\mathrm{max}$, where the maximum radius ${\rho}_\mathrm{max}$ takes the form
\begin{equation}
	{\rho}_{\mathrm{max}} = \frac{8}{27\tkb^2}.
\end{equation}
Parabolic generalized cardioids, unlike logarithmic spirals or Keplerian parabolas, are bounded (they never escape the gravitational attraction of the central body).

\begin{figure*}
	\centering
	\subfloat[Elliptic\label{Fig:cardioids_ellip}]{\includegraphics[width=.19\linewidth]{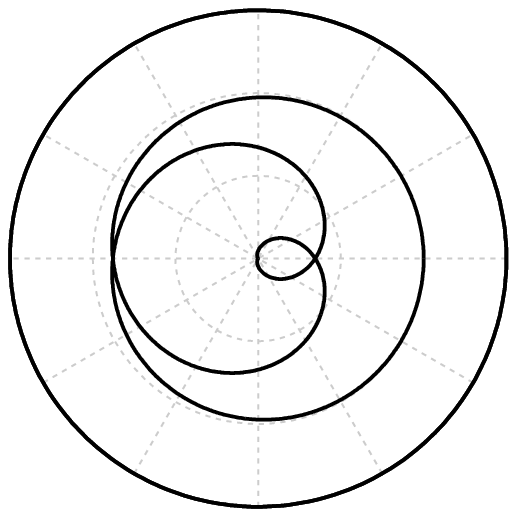}}\hspace{1mm}
	\subfloat[Parabolic\label{Fig:cardioids_parab}]{\includegraphics[width=.19\linewidth]{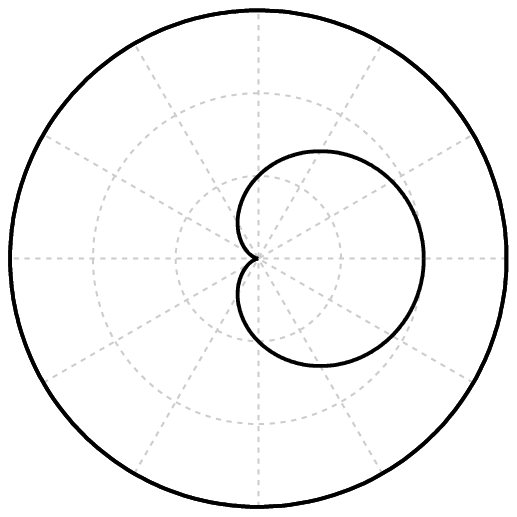}}\hspace{1mm}
	\subfloat[Hyperbolic Type I\label{Fig:cardioids_hypI}]{\includegraphics[width=.19\linewidth]{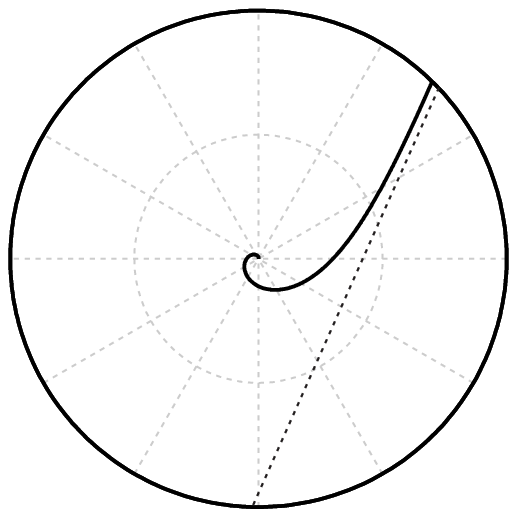}}\hspace{1mm}
	\subfloat[Hyperbolic Type II \label{Fig:cardioids_hypII}]{\includegraphics[width=.19\linewidth]{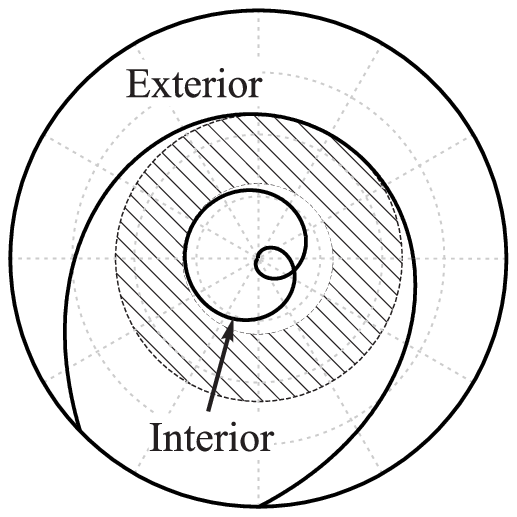}}\hspace{1mm}
	\subfloat[Hyperbolic transition\label{Fig:cardioids_trans}]{\includegraphics[width=.19\linewidth]{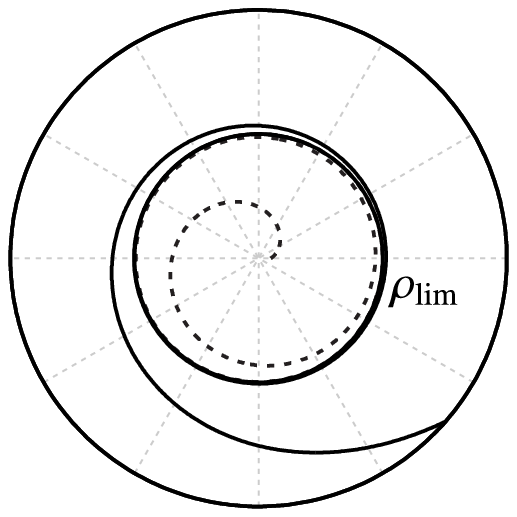}}
	\caption{Examples of generalized cardioids $(\gamma=3)$.}
\end{figure*}

The line of apses is defined by:
\begin{equation}
	\theta_m = \theta_0 + n\left[ \frac{\piup}{2} + \arcsin\left( 1 - \frac{27}{4}\tkb^2{\rho}_0 \right) \right].
\end{equation}
The equation of the trajectory reveals that the orbit is in fact a pure cardioid:\footnote{The cardioid is a particular case of the lima\c{c}ons, curves first studied by the amateur mathematician \'Etienne Pascal in the 17th century. The general form of the lima\c{c}on in polar coordinates is $r(\theta)=a+b\cos\theta$. Depending on the values of the coefficients the curve might reach the origin and form loops. It is worth noticing that the inverse of a lima\c{c}on, $r(\theta)=(a+b\cos\theta)^{-1}$, results in a conic section. Lima\c{c}ons with $a=b$ are considered part of the family of sinusoidal spirals.}
\begin{equation}\label{Eq:traj_parab_gamma3}
	\frac{{\rho}(\theta)}{{\rho}_\mathrm{max}} = \frac{1}{2}[1+\cos(\theta-\theta_m)].
\end{equation}
This curve is symmetric with respect to $\theta_m$. Figure~\ref{Fig:cardioids_parab} depicts the geometry of the solution.

\subsection{Hyperbolic motion}
The inequality in Eq.~\eqref{Eq:ineq} determines the structure of the solutions and the sign of the discriminant~\eqref{Eq:discriminant} governs the nature of its roots. There are two types of hyperbolic generalized cardioids: for $\tkb<\sqrt{3\tka}$ the cardioids are of Type I, and for $\tkb>\sqrt{3\tka}$ the cardioids are of Type II.

\subsubsection{Hyperbolic cardioids of Type I}
For hyperbolic cardioids of Type I there is only one real root. The other two are complex conjugates. The real root is
\begin{equation}\label{Eq:r1_hyperbI_g3}
	%{\rho}_1 = - 2\sqrt{\frac{3\tkb^2}{\tka^3}}\cosh \left[ \frac{1}{3}\arccosh\left(  \frac{\sqrt{3\tka}}{\tkb}   \right) \right]- \frac{2}{3\tka} 
	%
	%
	{\rho}_3 = -\frac{\Lambda\tka^2(2+\Lambda\tka)+3\tkb^2}{3\tka^3\Lambda},
\end{equation}
having introduced the auxiliary parameter
\begin{equation}
	\Lambda = \left\{  \frac{3\tkb^2}{\tka^{9/2}}\left[ 3\sqrt{\tka} + \sqrt{3(3\tka-\tkb^2)} \right]  \right\}^{1/3}.
\end{equation}
Provided that $\Lambda>0$, Eq.~\eqref{Eq:r1_hyperbI_g3} shows that ${\rho}_3<0$. Therefore, there are no limits to the values that $\rho$ can take. As a consequence, the cardioid never transitions between regimes. If it is initially $\psi_0<\piup/2$ it will always escape to infinity, and fall toward the origin for $\psi_0>\piup/2$.

The equation of the trajectory for hyperbolic generalized cardioids of Type I is
\begin{equation}\label{Eq:traj_hyperI_gamma3}
	\frac{{\rho}(\theta)}{{\rho}_3A} =   \frac{(A+B)\sn^2(\nu,k) + 2B[\cn(\nu,k)-1]}{(A+B)^2\sn^2(\nu,k)-4AB} .
\end{equation}
It is defined in terms of 
\begin{equation}
	A = \sqrt{b_1^2 + a_1^2}, \qquad B = \sqrt{  ({\rho}_3 - b_1)^2 + a_1^2  },
\end{equation}
which require
\begin{equation}
	b_1 = \frac{\Lambda\tka^2(\Lambda\tka-4)+3\tkb^2}{6\tka^3\Lambda},\quad a_1^2 = \frac{(\Lambda^2\tka^3-3\tkb^2)^2}{6\Lambda^2\tka^6}.
\end{equation}
There are no axes of symmetry. Therefore, the anomaly is referred directly to the initial conditions:
\begin{equation}
	\nu(\theta) = \frac{\tka^{3/2}}{\tkb}\sqrt{AB}\,(\theta-\theta_0) + \EllF(\phi_0,k).
\end{equation}
The moduli of both the Jacobi elliptic functions and the elliptic integral, and the argument of the latter, are
\begin{equation}
	k = \sqrt{\frac{(A+B)^2-{\rho}_3^2}{4AB}},\quad \phi_0 = \arccos\left[ \frac{A\lambda_{03}-B\rho_0}{A\lambda_{03}+B\rho_0} \right].
\end{equation}
The cardioid approaches infinity along an asymptotic branch, with $\tilde{v}\to\tilde{v}_\infty$ as ${\rho}\to\infty$. The orientation of the asymptote follows from the limit
\begin{equation}
	\theta_\mathrm{as} = \lim_{{\rho}\to\infty}\theta({\rho}) = \theta_0 + \frac{n\tkb}{\tka^{3/2}\sqrt{AB}}\big[\EllF(\phi_\infty,k) - \EllF(\phi_0,k)\big].
\end{equation}
Here, the value of $\phi_\infty$,
\begin{equation}
	\phi_\infty = \arccos\left( \frac{A-B}{A+B} \right),
\end{equation}
is defined as $\phi_\infty=\lim_{{\rho}\to\infty}\phi$. An example of a hyperbolic cardioid of Type I with its corresponding asymptote is presented in Fig.~\ref{Fig:cardioids_hypI}.

\subsubsection{Hyperbolic cardioids of Type II}
For hyperbolic cardioids of Type II the polynomial in Eq.~\eqref{Eq:ineq} admits three distinct real roots, $\{{\rho}_1,{\rho}_2,{\rho}_3\}$, given by
\begin{equation}
	{\rho}_{k+1} = \frac{2\tkb}{\sqrt{3\tka^3}}\cos\left[ \frac{\piup}{3}(2k+1) - \frac{1}{3}\arccos\left( \frac{\sqrt{3\tka}}{\tkb} \right) \right] - \frac{2}{3\tka} 
\end{equation}
with $k=0,1,2$. The roots are then sorted so that ${\rho}_1>{\rho}_2>{\rho}_3$. Since
\begin{equation}
	{\rho}_1{\rho}_2{\rho}_3 = -\frac{8}{27\tka^3}<0, 
\end{equation}
then ${\rho}_3<0$ and also ${\rho}_1>{\rho}_2>0$ for physical coherence. The polynomial constraint reads
\begin{equation}
	({\rho}-{\rho}_1)({\rho}-{\rho}_2)({\rho}-{\rho}_3)\geq0.
\end{equation}
and holds for both ${\rho}>{\rho}_1$ and ${\rho}<{\rho}_2$. The integral of motion~\eqref{Eq:integr1} shows that both situations are physically admissible, because $\tilde{v}^2(\rho_1)>0$ and $\tilde{v}^2(\rho_2)>0$. There are two families of solutions that lie outside the annulus ${\rho}\not\in[{\rho}_2,{\rho}_1]$. When ${\rho}_0\leq{\rho}_2$ the spirals are \emph{interior}, whereas for ${\rho}\geq{\rho}_1$ they are \emph{exterior} spirals. The geometry of the forbidden region can be analyzed in Fig.~\ref{Fig:cardioids_hypII}. The particle cannot enter the barred annulus, whose limits coincide with the periapsis and apoapsis of the exterior and interior orbits, respectively.

The axis of symmetry of interior spirals is given by 
\begin{equation}
	\theta_m = \theta_0 + \frac{2n\tkb [\EllK(k)-\EllF(\phi_0,k)]}{\tka^{3/2}\sqrt{{\rho}_1\lambda_{23}}}
\end{equation}
in terms of the arguments:
\begin{equation}
	\phi_0 = \arcsin\sqrt{\frac{{\rho}_0\lambda_{23}}{{\rho}_2\lambda_{03}}},\qquad k =\sqrt{ \frac{{\rho}_2\lambda_{13}}{{\rho}_1\lambda_{23}}}.
\end{equation}
The trajectory simplifies to:
\begin{equation}\label{Eq:traj_hypII_g3_int}
	 %{{\rho}(\theta)}  = \frac{ {\rho}_3 \cd^2(\nu,k)}{({\rho}_3/\rho_2) +   \cd^2(\nu,k)-1 }
	 \frac{{\rho}(\theta)}{\rho_3}  = \frac{ 1 }{1 + (\lambda_{32}/\rho_2)\dc^2(\nu,k)  }.
\end{equation}
Here we made use of Glaisher's notation for the Jacobi elliptic functions,\footnote{Glaisher's notation establishes that if p,q,r are any of the four letters s,c,d,n, then:
\begin{equation}
	\mathrm{pq}(\nu,k) = \frac{\mathrm{pr}(\nu,k)}{\mathrm{qr}(\nu,k)}=\frac{1}{\mathrm{qp}(\nu,k)}.
\end{equation}
Under this notation repeated letters yield unity.} so $\dc(\nu,k)=\dn(\nu,k)/\cn(\nu,k)$. The spiral anomaly $\nu$ takes the form
\begin{equation}
	\nu(\theta) = \frac{\tka^{3/2}\sqrt{{\rho}_1\lambda_{23}}}{2\tkb}\,(\theta-\theta_m).
\end{equation}
The trajectory is symmetric with respect to the line of apses defined by $\theta_m$.

For the case of exterior spirals the largest root $\rho_1$ behaves as the periapsis. A cardioid initially in lowering regime will reach ${\rho}_1$, then it will transition to raising regime and escape to infinity. Equation~\eqref{Eq:to_integrate} is integrated from the initial radius to the periapsis to provide the orientation of the line of apses:
\begin{equation}\label{Eq:thetam_gamma3_hypII}
	\theta_m  =\theta_0 - \frac{2n\tkb\,\EllF(\phi_0,k)}{\tka^{3/2}\sqrt{{\rho}_1\lambda_{23}}},
\end{equation}
with
\begin{equation}
	\phi_0 = \arcsin\sqrt{\frac{\lambda_{23}\lambda_{01}}{\lambda_{13}\lambda_{02}}},\quad k = \sqrt{\frac{{\rho}_2\lambda_{13}}{{\rho}_1\lambda_{23}}}
\end{equation}
the argument and modulus of the elliptic integral.

The trajectory of exterior spirals is obtained upon inversion of the equation for the polar angle,
\begin{equation}\label{Eq:traj_hypII_gamma3}
	{\rho}(\theta) = \frac{{\rho}_2\lambda_{13}\sn^2(\nu,k)-{\rho}_1\lambda_{23}}{\lambda_{13}\sn^2(\nu,k)-\lambda_{23}}.
\end{equation}
The anomaly is redefined as
\begin{equation}
	\nu(\theta) =  \frac{\tka^{3/2}\sqrt{{\rho}_1\lambda_{23}}}{2\tkb}\,(\theta-\theta_m).
\end{equation}
This variable is referred to the line of apses, given in Eq.~\eqref{Eq:thetam_gamma3_hypII}. The form of the solution shows that hyperbolic generalized cardioids of Type II are symmetric with respect to $\theta_m$.

Due to the symmetry of Eq.~\eqref{Eq:traj_hypII_gamma3} the trajectory exhibits two symmetric asymptotes, defined by
\begin{equation}
	\theta_\mathrm{as}  =\theta_m \pm \frac{2\tkb\EllF(\phi_\infty,k)}{\tka^{3/2}\sqrt{{\rho}_1\lambda_{23}}}.
\end{equation}
The argument $\phi_\infty$ reads
\begin{equation}
	\phi_\infty = \arcsin\sqrt{\frac{\lambda_{23}}{ \lambda_{13} }}.
\end{equation}

\subsubsection{Transition between Type I and Type II hyperbolic cardioids}
The limit case $\tkb=\sqrt{3\tka}$ defines the transition between hyperbolic cardioids of Types I and II. The discriminant $\Delta$ vanishes: the roots are all real and one is a multiple root, $\rho_\mathrm{lim}\equiv{\rho}_1={\rho}_2$. The region of forbidden motion degenerates into a circumference of radius $\rho_\mathrm{lim}$. The roots take the form:
\begin{equation}\label{Eq:lim_radius_cardioid_Trans}
	{\rho}_{3} =   - \frac{8}{3\tka}<0 \qquad \text{and}\qquad {\rho}_\mathrm{lim}\equiv{\rho}_1={\rho}_2 = \frac{1}{3\tka}.
\end{equation}
The condition $({\rho}-{\rho}_3)({\rho}-{\rho}_\mathrm{lim})^2 \geq0$ holds naturally for ${\rho}>0$. When the cardioid reaches ${\rho}_\mathrm{lim}$ the velocity becomes $\tilde{v}_\mathrm{lim} = \sqrt{3\tka} = 1/{\sqrt{{\rho}_\mathrm{lim}}}$. It coincides with the local circular velocity. Moreover, from the integral of motion~\eqref{Eq:integr2} one has
\begin{equation}
	\tkb = {\rho}_\mathrm{lim}\tilde{v}_\mathrm{lim}^3\sin\psi_\mathrm{lim} \implies \sin\psi_\mathrm{lim} = 1
\end{equation}
meaning that the orbit becomes circular as the particle approaches ${\rho}_\mathrm{lim}$. When $\psi_\mathrm{lim}=\piup/2$ the perturbing acceleration in Eq.~\eqref{Eq:ap_transformed} vanishes. As a result, the orbit degenerates into a circular Keplerian orbit. A cardioid with ${\rho}_0<{\rho}_\mathrm{lim}$ and in raising regime will reach ${\rho}_\mathrm{lim}$ and degenerate into a circular orbit with radius ${\rho}_\mathrm{lim}$. This phenomenon also appears in cardioids with ${\rho}_0>{\rho}_\mathrm{lim}$ and in lowering regime.

The trajectory reduces to
\begin{equation}\label{Eq:traj_gamma3_hypertrans}
	\frac{{\rho}(\theta)}{{\rho}_\mathrm{lim}} = \frac{\cosh\nu-1}{\cosh\nu+5/4},
\end{equation}
which is written in terms of the anomaly
\begin{equation}
	\nu(\theta) = \frac{\theta-\theta_0}{\sqrt{3}} + 2mn\arctanh\left( 3\sqrt{ \frac{ {\rho}_0}{8{\rho}_\mathrm{lim}+ {\rho}_0} } \right).
\end{equation}
The integer $m=\sign(1-{\rho}_0/{\rho}_\mathrm{lim})$ determines whether the particle is initially below $(m=+1)$ or above $(m=-1)$ the limit radius ${\rho}_\mathrm{lim}$. The limit $\lim_{\theta\to\infty}{\rho}  = {\rho}_\mathrm{lim} = 1/({3\tka})$ shows that the radius converges to ${\rho}_\mathrm{lim}$. This limit only applies to the cases $m=n=+1$ and $m=n=-1$. When the particle is initially below ${\rho}_\mathrm{lim}$ and in lowering regime, $\{m=+1,n=-1\}$, it falls toward the origin. In the opposite case, $\{m=-1,n=+1\}$, the cardioid approaches infinity along an asymptotic branch with
\begin{equation}
	\theta_\mathrm{as} = \theta_0 - 2\sqrt{3}\left[mn\arctanh\left( 3\sqrt{ \frac{ {\rho}_0}{8{\rho}_\mathrm{lim}+ {\rho}_0} } \right) + \arctanh(3)\right].
\end{equation}
Two example trajectories with $n=+1$ are plotted in Fig.~\ref{Fig:cardioids_trans}. The dashed line corresponds to $m=+1$ and the solid line to $m=-1$. The trajectories terminate/emanate from a circular orbit of radius $\rho_\mathrm{lim}$.

\section{Case $\gamma=4$: generalized sinusoidal spirals}\label{Sec:sinusoidal}
Setting $\gamma=4$ in Eq.~\eqref{Eq:ineq} gives rise to the polynomial inequality
%\begin{equation}\label{Eq:ineq_gamma4}
%	  \tka^2{\rho}^2 + (\tka - \tkb){\rho} + \frac{1}{4}\geq0,
%\end{equation}
\begin{equation}\label{Eq:ineq_gamma4}
	 %\left[  (\tka \rho + 1)\tka \rho + \frac{1}{4} + \tkb\rho \right]\left[  (\tka \rho + 1)\tka \rho + \frac{1}{4} - \tkb\rho \right] \geq0,
	 %
	 \left[  4(\tka^2 \rho + \tka + \tkb) \rho + 1 \right]\left[  4(\tka^2 \rho + \tka - \tkb) \rho + 1 \right] \geq0,
\end{equation}
which governs the subfamilies of the solutions to the problem. The four roots of the polynomial are
\begin{equation}
	\begin{split}
	 {\rho}_{1,2} &= +\frac{\tkb-\tka\pm\sqrt{\tkb(\tkb-2\tka)}}{2\tka^2} \label{Eq:roots_ellip_gamma4} \\
	 {\rho}_{3,4} &= -\frac{\tka+\tkb\mp\sqrt{\tkb(\tkb+2\tka)}}{2\tka^2}
	\end{split}
\end{equation}
and the discriminant of $P_\mathrm{nat}(\rho)$ is 
\begin{equation}\label{Eq:discriminant_g4}
	\Delta = \frac{\tkb^6}{\tka^{20}}(\tkb^2-4\tka^2).
\end{equation}
The sign of the discriminant determines the nature of the four roots. 

%{Eq:dthetadr_integrability}

\subsection{Elliptic motion}
When $\tka<0$ there are two subfamilies of elliptic sinusoidal spirals: of Type I, with $\tkb>-2\tka$ ($\Delta>0$), and of Type II, with $\tkb<-2\tka$ ($\Delta<0$). Both types are separated by the limit case $\tkb=-2\tka$, which makes $\Delta=0$.

\subsubsection{Elliptic sinusoidal spirals of Type I}
For the case $\tkb>-2\tka$ the discriminant is positive and the four roots are real, with $\rho_1>\rho_2>\rho_3>\rho_4$. Since $\rho_{3,4}<0$ Eq.~\eqref{Eq:ineq_gamma4} reduces to
\begin{equation}\label{Eq:condition_binomials}
	(\rho-\rho_1)(\rho-\rho_2)\geq0,
\end{equation}
meaning that it must be either $\rho>\rho_1$ or $\rho<\rho_2$. The integral of motion~\eqref{Eq:integr1} reveals that only the latter case is physically possible, because $\tilde{v}^2(\rho_1)<0$. Thus, $\rho_2$ is the apoapsis of the spiral:
\begin{equation}
	\rho_\mathrm{max} \equiv \rho_2 = \frac{\tkb-\tka - \sqrt{\tkb(\tkb-2\tka)}}{2\tka^2}
\end{equation}
and $\rho\leq\rho_\mathrm{max}$. Equation~\eqref{Eq:dthetadr_integrability} is then integrated from ${\rho}_0$ to ${\rho}_\mathrm{max}$ to define the orientation of the apoapsis,
\begin{equation}
	\theta_m = \theta_0 - \frac{2n\tkb[\EllF(\phi_0,k)-\EllK(k)]}{\tka^2\sqrt{\lambda_{13}\lambda_{24}}}.
\end{equation}
The arguments of the elliptic integrals are
\begin{equation}
	\phi_0 = \arcsin\sqrt{\frac{\lambda_{24}\lambda_{03}}{\lambda_{23}\lambda_{04}}},\quad k = \sqrt{\frac{\lambda_{23}\lambda_{14}}{\lambda_{13}\lambda_{24}}}.
\end{equation}	
Recall that $\lambda_{ij}=\rho_i-\rho_j$.

Elliptic sinusoidal spirals of Type I are defined by 
\begin{equation}\label{Eq:traj_ellipI_gamma4}
	 {\rho}(\theta) = \frac{{\rho}_4\lambda_{23}\cd^2(\nu,k) - {\rho}_3\lambda_{24}}{\lambda_{23}\cd^2(\nu,k)- \lambda_{24}}.
\end{equation}
The spiral anomaly reads
\begin{equation}
	\nu(\theta) = \frac{n\tka^2}{2\tkb}(\theta-\theta_m)\sqrt{\lambda_{13}\lambda_{24}} .
\end{equation}
The trajectory is symmetric with respect to $\theta_m$, which corresponds to the line of apses.

\subsubsection{Elliptic sinusoidal spirals of Type II}
When $\tkb<-2\tka$ two roots are real and the other two are complex conjugates. In this case the real roots are $\rho_{1,2}$ and the inequality~\eqref{Eq:ineq_gamma4} reduces again to Eq.~\eqref{Eq:condition_binomials}, meaning that $\rho\leq\rho_2\equiv\rho_\mathrm{max}$. However, the form of the solution is different from the trajectory described by Eq.~\eqref{Eq:traj_ellipI_gamma4}, being 
\begin{equation}\label{Eq:traj_ellipII_gamma4}
	{\rho}(\theta) = \frac{{\rho}_1B-{\rho}_2A - ({\rho}_2A+{\rho}_1B)\cn(\nu,k)}{B-A - (A+B)\cn(\nu,k)}.
\end{equation}
The spiral anomaly is
\begin{equation}
	\nu(\theta) = \frac{n\tka^2}{\tkb}\sqrt{AB}\,(\theta-\theta_m)  .
\end{equation}
It is referred to the orientation of the apse line, $\theta_m$. This variable is given by
\begin{equation}
	\theta_m = \theta_0 + \frac{n\tkb}{\tka^2\sqrt{AB}}[2\EllK(k)-\EllF(\phi_0,k)]
\end{equation}
considering the arguments:
\begin{alignat}{2}
	 \phi_0 = \arccos\left[ \frac{A\lambda_{02} + B\lambda_{10}}{A\lambda_{02} - B\lambda_{10}} \right], \qquad k = \sqrt{\frac{(A+B)^2-\lambda_{12}^2}{4AB}}.
\end{alignat}
The coefficients $A$ and $B$ are defined in terms of 
\begin{equation}
	a_1^2 = -\frac{\tkb(\tkb+2\tka)}{4\tka^4}\quad\text{and}\quad b_1 = -\frac{\tka+\tkb}{2\tka^2},
\end{equation}
namely
\begin{equation}
	A = \sqrt{({\rho}_1-b_1)^2+a_1^2},\quad B = \sqrt{({\rho}_2-b_1)^2+a_1^2}.
\end{equation}
The fact that the Jacobi function $\cn(\nu,k)$ is symmetric proves that elliptic sinusoidal spirals of Type II are symmetric.

\subsubsection{Transition between spirals of Types I and II}
In this is particular case of elliptic motion, $-2\tka=\tkb$, the roots of polynomial $P_\mathrm{nat}({\rho})$ are 
\begin{equation}
	{\rho}_{1} = \frac{3+2\sqrt{2}}{\tkb}, \quad {\rho}_2 \equiv \rho_\mathrm{max}=\frac{3-2\sqrt{2}}{\tkb},\quad {\rho}_{3,4} = -\frac{1}{\tkb}.
\end{equation}
These results simplify the definition of the line of apses to
\begin{equation}
	\theta_m  =\theta_0 +  \sqrt{2}\,n\ln\left[ \frac{\sqrt{2}(1-{\rho}_0\tkb)+\sqrt{(3-{\rho}_0\tkb)^2-8}}{1+{\rho}_0\tkb} \right].
\end{equation}	
Introducing the spiral anomaly $\nu(\theta)$,
\begin{equation}
	\nu(\theta) = \frac{\sqrt{2}}{2}(\theta-\theta_m),
\end{equation}
the equation of the trajectory takes the form
\begin{equation}\label{Eq:traj_elliplim_gamma4}
	\frac{{\rho}(\theta)}{{\rho}_\mathrm{max}} = \frac{5-4\sqrt{2}\cosh\nu+\cosh(2\nu)}{(3-2\sqrt{2})[3-\cosh(2\nu)]}.
\end{equation}
The trajectory is symmetric with respect to the line of apses $\theta_m$ thanks to the symmetry of the hyperbolic cosine. 

Figure~\ref{Fig:sinusoidal_ellip} shows the three types of elliptic spirals. It is important to note that in all three cases the condition in Eq.~\eqref{Eq:ineq_gamma4} transforms into Eq.~\eqref{Eq:condition_binomials}, equivalent to $\rho<\rho_\mathrm{max}$. As a result, there are no differences in their nature although the equations for the trajectory are different.

\subsection{Parabolic motion: sinusoidal spiral (off-center circle)}
Making $\tka=0$, the condition in Eq.~\eqref{Eq:ineq_gamma4} simplifies to
\begin{equation}
	{\rho}\leq {\rho}_\mathrm{max} = \frac{1}{4\tkb},
\end{equation}
meaning that the spiral is bounded by a maximum radius ${\rho}_\mathrm{max}$. It is equivalent to the apoapsis of the spiral. Its orientation is given by
\begin{equation}
	\theta_m = \theta_0 + n\left[\frac{\piup}{2} - \arcsin\left(\frac{{\rho}_0}{{\rho}_\mathrm{max}}\right)  \right].
\end{equation}

\begin{figure*}
	\centering
	\subfloat[Elliptic\label{Fig:sinusoidal_ellip}]{\includegraphics[width=.19\linewidth]{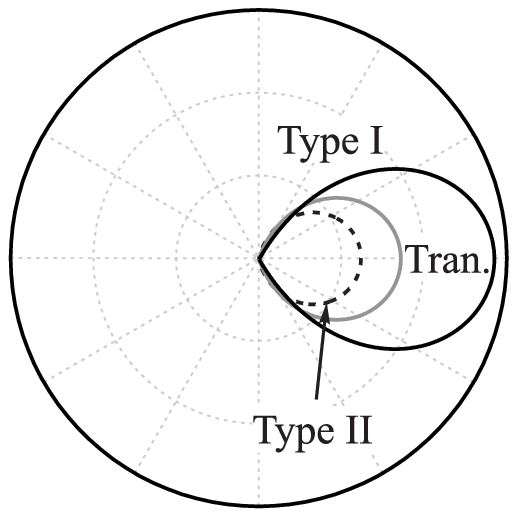}}\hspace{1mm}
	\subfloat[Parabolic\label{Fig:sinusoidal_parab}]{\includegraphics[width=.19\linewidth]{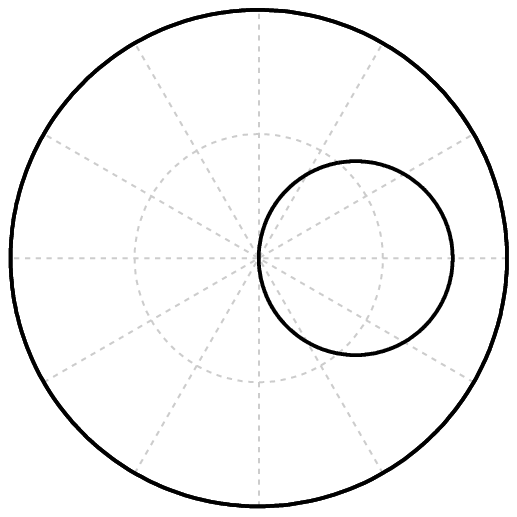}}\hspace{1mm}
	\subfloat[Hyperbolic Type I\label{Fig:sinusoidal_hypI}]{\includegraphics[width=.19\linewidth]{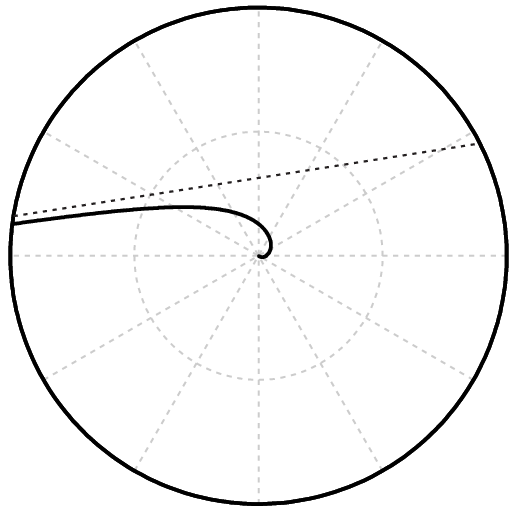}}\hspace{1mm}
	\subfloat[Hyperbolic Type II\label{Fig:sinusoidal_hypII}]{\includegraphics[width=.19\linewidth]{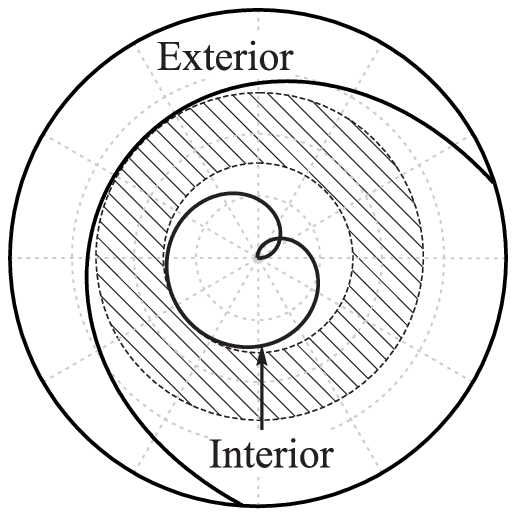}}\hspace{1mm}
	\subfloat[Hyperbolic transition\label{Fig:sinusoidal_trans}]{\includegraphics[width=.19\linewidth]{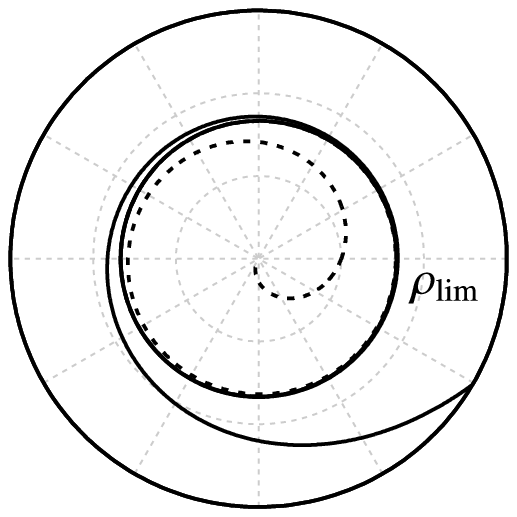}}
	\caption{Examples of generalized sinusoidal spirals $(\gamma=4)$.}
\end{figure*}

The trajectory reduces to a sinusoidal spiral,\footnote{It was the Scottish mathematician Colin Maclaurin the first to study sinusoidal spirals. In his ``{Tractatus de Curvarum Constructione \& Mensura}'', published in \emph{Philosophical Transactions} in 1717, he constructed this family of curves relying on the epicycloid. Their general form is $r^n=\cos(n\theta)$ and different values of $n$ render different types of curves; $n=-2$ correspond to hyperbolas, $n=-1$ to straight lines, $n=-1/2$ to parabolas, $n=-1/3$ to Tschirnhausen cubics, $n=1/3$ to Cayley's sextics, $n=1/2$ to cardioids, $n=1$ to circles, and $n=2$ to lemniscates.} and its definition can be directly related to $\theta_m$:
\begin{equation}\label{Eq:traj_parab_gamma4}
	\frac{{\rho}(\theta)}{{\rho}_\mathrm{max}} = \cos(\theta - \theta_m).
\end{equation}
The spiral defined in Eq.~\eqref{Eq:traj_parab_gamma4} is symmetric with respect to the line of apses. The resulting orbit is a circle centered at $({\rho}_\mathrm{max}/2,\theta_m)$ (Fig.~\ref{Fig:sinusoidal_parab}). Circles are indeed a special case of sinusoidal spirals.

\subsection{Hyperbolic motion}
Given the discriminant in Eq.~\eqref{Eq:discriminant_g4}, for the case $\tka>0$ the values of the constant $\tkb$ define two different types of hyperbolic sinusoidal spirals: spirals of Type I ($\tkb<2\tka$), and spirals of Type II ($\tkb>2\tka$). 

%For spirals of Type I the discriminant described in Eq.~\eqref{Eq:discriminant_g4} is negative. The quartic appearing in Eq.~\eqref{Eq:dthetadr_integrability} has two real and two complex roots. Conversely, the four roots are real in the case of spirals of Type II.

\subsubsection{Hyperbolic sinusoidal spirals of Type I}
If $\tkb<2\tka$, then $\rho_{1,2}$ are complex conjugates and $\rho_{3,4}$ are both real but negative. Therefore, the condition in Eq.~\eqref{Eq:ineq_gamma4} holds naturally for any radius and there are no limitations to the values of $\rho$. The particle can either fall to the origin or escape to infinity along an asymptotic branch. The equation of the trajectory is given by:
\begin{equation}\label{Eq:traj_hyperI_gamma4}
	{\rho}(\theta) = \frac{{\rho}_3B - {\rho}_4A + ({\rho}_3B + {\rho}_4A)\cn(\nu,k)}{B-A + (A+B)\cn(\nu,k)},
\end{equation}
where the spiral anomaly can be referred directly to the initial conditions:
\begin{equation}
	\nu(\theta) = \frac{n\tka^2}{\tkb}\sqrt{AB}\,(\theta-\theta_0) + \EllF(\phi_0,k).
\end{equation}
This definition involves an elliptic integral of the first kind with argument and parameter:
\begin{alignat}{2}
	  \phi_0 = \arccos\left[ \frac{A\lambda_{04}+B\lambda_{30}}{A\lambda_{04}-B\lambda_{30}} \right],\qquad k  = \sqrt{\frac{(A+B)^2-\lambda_{34}^2}{4AB}}.
\end{alignat}
The coefficients $A$ and $B$ require the terms:
\begin{equation}
	b_1 = -\frac{\tka+\tkb}{2\tka^2},\qquad a_1^2 = -\frac{\tkb(\tkb+2\tka)}{4\tka^4},
\end{equation}
being
\begin{equation}
	A = \sqrt{({\rho}_3-b_1)^2+a_1^2},\quad B = \sqrt{({\rho}_4-b_1)^2+a_1^2}.
\end{equation}
The direction of the asymptote is defined by
\begin{equation}
	\theta_\mathrm{as} = \theta_0 + \frac{n\tkb}{\tka^2\sqrt{AB}}\left[ \EllF(\phi_\infty,k) - \EllF(\phi_0,k) \right].
\end{equation}
This definition involves the argument
\begin{equation}
	\phi_\infty = \arccos\left( \frac{A-B}{A+B} \right).
\end{equation}
The velocity of the particle when reaching infinity is $\tilde{v}_\infty=\sqrt{\tka}$. Hyperbolic sinusoidal spirals of Type I are similar to the hyperbolic solutions of Type I with $\gamma=2$ and $\gamma=3$. Figure~\ref{Fig:sinusoidal_hypII} depicts and example trajectory and the asymptote defined by $\theta_\mathrm{as}$.

%\begin{equation}
%	\tkc = \theta_0 - \frac{n\tkb}{\tka^2}g\EllF(\phi_0,k)
%\end{equation}

\subsubsection{Hyperbolic sinusoidal spirals of Type II}
In this case the four roots are real and distinct, with $\rho_{3,4}<0$. The two positive roots $\rho_{1,2}$ are physically valid, i.e. $\tilde{v}^2({\rho}_1)>0$ and $\tilde{v}^2({\rho}_2)>0$. This yields two situations in which the condition $({\rho}-{\rho}_1)({\rho}-{\rho}_2)\geq0$ is satisfied: ${\rho}>{\rho}_1$ (exterior spirals) and ${\rho}<{\rho}_2$ (interior spirals).

Interior spirals take the form
\begin{equation}\label{Eq:traj_hyperIIint_gamma4}
	{\rho}(\theta) = \frac{{\rho}_2\lambda_{13} - {\rho}_1\lambda_{23}\sn^2(\nu,k)}{\lambda_{13} - \lambda_{23}\sn^2(\nu,k)}.
\end{equation}
The spiral anomaly is 
\begin{equation}
	\nu(\theta) = \frac{\tka^2}{2\tkb}\sqrt{\lambda_{13}\lambda_{24}}\,(\theta-\theta_m).
\end{equation}
The orientation of the line of apses is solved from
\begin{equation}\label{Eq:thetam_hyperbII_gamma4}
	\theta_m = \theta_0 + \frac{2n\tkb\EllF(\phi_0,k)}{\tka^2\sqrt{\lambda_{13}\lambda_{24}}},
\end{equation}
with
\begin{equation}
	\phi_0 = \arcsin\sqrt{\frac{\lambda_{13}\lambda_{20}}{\lambda_{23}\lambda_{10}}},\qquad k = \sqrt{\frac{\lambda_{23}\lambda_{14}}{\lambda_{13}\lambda_{24}}}.
\end{equation}
Interior hyperbolic spirals are bounded and their shape is similar to that of a lima\c{c}on.

The line of apses of an exterior spiral is defined by
\begin{equation}
	\theta_m = \theta_0 - \frac{2n\tkb\EllF(\phi_0,k)}{\tka^2\sqrt{\lambda_{13}\lambda_{24}}}.
\end{equation}
The modulus and the argument of the elliptic integral are
\begin{equation}
	\phi_0 = \arcsin\sqrt{\frac{\lambda_{24}\lambda_{01}}{\lambda_{14}\lambda_{02}}},\qquad k = \sqrt{\frac{\lambda_{23}\lambda_{14}}{\lambda_{13}\lambda_{24}}}.
\end{equation}
The trajectory becomes
\begin{equation}\label{Eq:traj_hyperIIext_gamma4}
	{\rho}(\theta) = \frac{{\rho}_1\lambda_{24} + {\rho}_2\lambda_{41}\sn^2(\nu,k)}{\lambda_{24} + \lambda_{41}\sn^2(\nu,k)}
\end{equation}
and it is symmetric with respect to $\theta_m$. The geometry of the solution is similar to that of hyperbolic cardioids of Type II, mainly because of the existence of the forbidden region plotted in Fig.~\ref{Fig:sinusoidal_hypII}. The asymptotes follow the direction of
\begin{equation}
	\theta_\mathrm{as} = \theta_m \pm \frac{2\tkb\EllF(\phi_\infty,k)}{\tka^2\sqrt{\lambda_{13}\lambda_{24}}},\quad \text{with}\quad \phi_\infty = \arcsin\sqrt{\frac{\lambda_{24}}{\lambda_{14}}}.
\end{equation}

\subsubsection{Transition between Type I and Type II spirals}
When $\tkb=2\tka$ the radii ${\rho}_1$ and ${\rho}_2$ coincide, ${\rho}_1={\rho}_2\equiv\rho_{\mathrm{lim}}$, and become equal to the limit radius
\begin{equation}
	{\rho}_\mathrm{lim} = \frac{1}{2\tka}.
\end{equation}
The equation of the trajectory is a particular case of Eq.~\eqref{Eq:traj_hyperIIext_gamma4}, obtained with $\tkb\to2\tka$:
\begin{equation}\label{Eq:traj_hyperlim_gamma4}
	\frac{{\rho}(\theta)}{{\rho}_{\mathrm{lim}}} = \left[1 - \frac{8}{4+m(\sinh\nu-3\cosh\nu)}\right]^{-1}.
\end{equation}
The spiral anomaly can be referred to the initial conditions,
\begin{equation}
	\nu(\theta) = \frac{nm}{\sqrt{2}}(\theta-\theta_0)+\ln\left[ \frac{2(1+2\tka{\rho}_0)+\sqrt{2+8\tka{\rho}_0(3+\tka{\rho}_0)}}{m(1-2\tka{\rho}_0)} \right],
\end{equation}
avoiding additional parameters. Here $m=\sign(1-{\rho}_0/{\rho}_\mathrm{lim})$ determines whether the spiral is below or over the limit radius ${\rho}_\mathrm{lim}$. The asymptote follows from
\begin{equation}
	\theta_\mathrm{as} = \theta_0 -n\sqrt{2}\log(1-\sqrt{2}/2).
\end{equation}

Like in the case of hyperbolic cardioids, for $m=n=-1$ and $m=n=+1$ spirals of this type approach the circular orbit of radius ${\rho}_\mathrm{lim}$ asymptotically, i.e. $\lim_{\theta\to\infty}{\rho} = {\rho}_\mathrm{lim}$. When approaching a circular orbit the perturbing acceleration vanishes and the spiral converges to a Keplerian orbit. See Fig.~\ref{Fig:sinusoidal_trans} for examples of hyperbolic sinusoidal spirals with $\{m=+1,n=+1\}$ (dashed) and $\{m=-1,n=+1\}$ (solid).

\section{Summary}\label{Sec:summary}
The solutions presented in the previous sections are summarized in Table~\ref{Tab:summary}, organized in terms of the values of $\gamma$. Each family is then divided in elliptic, parabolic, and hyperbolic orbits. The table includes references to the corresponding equations of the trajectories for convenience. The orbits are said to be bounded if the particle can never reach infinity, because $r<r_\mathrm{lim}$.

\begin{table*}
	\caption{Summary of the families of solutions.\label{Tab:summary}}
	\begin{tabular*}{\linewidth}{@{\extracolsep{\fill}}ccccccc@{}}
		\hline \noalign{\smallskip}
		Family & Type & $\gamma$ & $\tka$ & $\tkb$ & Bounded & Trajectory \\
		\noalign{\smallskip}\hline\noalign{\smallskip}
		       & Elliptic   & 1 & $<0$ & $>\sqrt{-1/\tka}$& Y  & Eq.~\eqref{Eq:traj_Kepler} \\
		Conic sections & Parabolic  & 1 & $=0$  & $-$& N& Eq.~\eqref{Eq:traj_Kepler} \\
		       & Hyperbolic & 1 & $>0$ & $-$& N& Eq.~\eqref{Eq:traj_Kepler} \\
		\noalign{\smallskip}\hline\noalign{\smallskip}
		            & Elliptic    & 2 & $<0$ & $<1$& Y& Eq.~\eqref{Eq:traj_ellip_gamma2} \\
		Generalized & Parabolic   & 2 & $=0$ & $\leq1$& N& Eq.~\eqref{Eq:traj_parab_gamma2}$^\ast$ \\
		logarithmic & Hyperbolic T-I & 2 & $>0$ & $<1$& N& Eq.~\eqref{Eq:traj_hyperI_gamma2} \\
		spirals     & Hyperbolic T-II& 2 & $>0$ & $>1$& N& Eq.~\eqref{Eq:traj_hypII} \\
		            & Hyperbolic trans.& 2 & $>0$ & $=1$& N& Eq.~\eqref{Eq:traj_hyperlim_gamma2} \\
		\noalign{\smallskip}\hline\noalign{\smallskip}
		%\multirow{5}{*}{\noindent\begin{minipage}{1.3cm}\centering{Generalized cardioids}\end{minipage}}   %
		         & Elliptic    & 3 & $<0$ & $<1$           & Y& Eq.~\eqref{Eq:ellip_g3} \\
		         & Parabolic   & 3 & $=0$ & $\leq1$        & Y& Eq.~\eqref{Eq:traj_parab_gamma3}$^\dagger$ \\
		  Generalized       & Hyperbolic T-I & 3 & $>0$ & $<\sqrt{3\tka}$& N& Eq.~\eqref{Eq:traj_hyperI_gamma3} \\
		  cardioids       & Hyperbolic T-II (int)& 3 & $>0$ & $>\sqrt{3\tka}$& Y& Eq.~\eqref{Eq:traj_hypII_g3_int} \\
		         & Hyperbolic T-II (ext)& 3 & $>0$ & $>\sqrt{3\tka}$& N& Eq.~\eqref{Eq:traj_hypII_gamma3} \\
		         & Hyperbolic trans.& 3 & $>0$ & $=\sqrt{3\tka}$& Y/N& Eq.~\eqref{Eq:traj_gamma3_hypertrans} \\
		\noalign{\smallskip}\hline\noalign{\smallskip}
		                & Elliptic T-I         & 4 & $<0$ & $>-2\tka$ & Y& Eq.~\eqref{Eq:traj_ellipI_gamma4}\\
		                & Elliptic T-II        & 4 & $<0$ & $<-2\tka$ & Y& Eq.~\eqref{Eq:traj_ellipII_gamma4}      \\
		    Generalized & Elliptic trans.        & 4 & $<0$ & $=-2\tka$ & Y& Eq.~\eqref{Eq:traj_elliplim_gamma4} \\
		    	sinusoidal  & Parabolic          & 4 & $=0$ & $-$& Y& Eq.~\eqref{Eq:traj_parab_gamma4}$^\ddagger$ \\
		    	spirals     & Hyperbolic T-I        & 4 & $>0$ & $<2\tka$  & N& Eq.~\eqref{Eq:traj_hyperI_gamma4}  \\
		    	            & Hyperbolic T-II (int) & 4 & $>0$ & $>2\tka$  & Y& Eq.~\eqref{Eq:traj_hyperIIint_gamma4} \\
		    	            & Hyperbolic T-II (ext) & 4 & $>0$ & $>2\tka$  & N& Eq.~\eqref{Eq:traj_hyperIIext_gamma4} \\
		    	            & Hyperbolic trans.       & 4 & $>0$ & $=2\tka$  & Y/N& Eq.~\eqref{Eq:traj_hyperlim_gamma4} \\
		\noalign{\smallskip}\hline \noalign{\smallskip}
	\end{tabular*}

	\vspace{-2mm}	
	
	\begin{flushleft}
	{\small$^\ast$ Logarithmic spiral.\\
	$^\dagger$ Cardioid.\\
	$^\ddagger$ Sinusoidal spiral (off-center circle).}
\end{flushleft}		
\end{table*}

\section{Unified solution in Weierstrassian formalism}\label{Sec:Weierstrass}
The orbits can be unified introducing the Weierstrass elliptic functions. Indeed, Eq.~\eqref{Eq:dthetadr_integrability} furnishes the integral expression
\begin{equation}\label{Eq:integral_weier}
	\theta(r) - \theta_0 = \int_{r_0}^r \frac{k\,\mathrm{d}s}{\sqrt{f(s)}}
\end{equation}
with
\begin{equation}\label{Eq:quartic_fs}
	f(s)\equiv P_\mathrm{sol}(s) = a_0s^4 + 4a_1 s^3 + 6a_2 s^2 + 4a_3 s + a_4
\end{equation}
and $a_{0,1}\neq0$. Introducing the auxiliary parameters 
\begin{equation}
	\vartheta = f^\prime(\rho_0)/4 \qquad\text{and}\qquad  \varphi = f^{\prime\prime}(\rho_0)/24
\end{equation}
Eq.~\eqref{Eq:integral_weier} can be inverted to provide the equation of the trajectory \citep[][p.~454]{whittaker1927course},
\begin{align}{2}
	\rho(\theta) =  & \rho_0 + \frac{1}{2[\wp(z)-\varphi]^2-f(\rho_0)f^{\mathrm{(iv)}}(\rho_0)/48} \nonumber\\[3mm]
	&    \times \Big\{ [\wp(z)-\varphi]2\vartheta - \wp^\prime(z)  \sqrt{f(\rho_0)}  + f(\rho_0)f^{\prime\prime\prime}(\rho_0)/24  \Big\}. \label{Eq:solution_weier}
\end{align}
The solution is written in terms of the Weierstrass elliptic function
\begin{equation}
	\wp(z)\equiv \wp(z;g_2,g_3)
\end{equation}
and its derivative $\wp^\prime(z)$, where $z=(\theta-\theta_0)/k$ is the argument and the invariant lattices $g_2$ and $g_3$ read
\begin{equation}
\begin{split}
	 g_2 & = a_0a_4-4a_1a_3+3a_2^2 \\
	 g_3 & = a_0a_2a_4 + 2a_1a_2a_3 - a_2^3 - a_0a_3^2 - a_1^2a_4.
\end{split}
\end{equation}
The coefficients $a_i$ and $k$ are obtained by identifying Eq.~\eqref{Eq:integral_weier} with Eq.~\eqref{Eq:dthetadr_integrability} for different values of $\gamma$. They can be found in Table~\ref{Tab:coeffs_weierstrass}. 
\begin{table}
	\caption{Coefficients $a_i$ of the polynomial $f(s)$, and factor $k$.\label{Tab:coeffs_weierstrass}}
	\begin{tabular*}{\linewidth}{@{\extracolsep{\fill}}lrrrrrr@{}}
	\hline \noalign{\smallskip}
		$\gamma$ & $a_0$ & $a_1$ & $a_2$ & $a_3$ & $a_4$ & $k$  \\
		\noalign{\smallskip}\hline\noalign{\smallskip}
		1 & $\tka$ & 1/2 & $-\tkb^2/6$ & 0 & 0 & $n\tkb$\\
		2 & $\tka^2$ & $\ka/2$ & $(1-\tkb^2)/6$ & 0 & 0 & $n\tkb$ \\
		3 & $27\tka^3$ & $27\tka^2/2$ & $6\tka - 9\tkb^2/2$ & 2 & 0 & $\sqrt{27}\,n\tkb$\\
		4 & $16\tka^4$ & $8\tka^3$ & $4\tka^2 - 8\tkb^2/3$ & $2\tka$ & $1$ & $4n\tkb$\\
		\noalign{\smallskip}\hline 
	\end{tabular*}
\end{table}

Symmetric spirals reach a minimum or maximum radius $\rho_m$, which is a root of $f(\rho)$. Thus, Eq.~\eqref{Eq:solution_weier} can be simplified if referred to $\rho_m$ instead of $\rho_0$:
\begin{equation}\label{Eq:solution_weier_symmetric}
	\rho(\theta) - \rho_m = \frac{f^\prime(\rho_m)/4}{\wp(z_m)-f^{\prime\prime}(z_m)/24}.
\end{equation}
This is the unified solution for all symmetric solutions, with $z_m=(\theta-\theta_m)/k$. Practical comments on the implementation of the Weierstrass elliptic functions can be found in \citet{biscani2014stark}, for example. Although $\wp(z)=\wp(-z)$, the derivative $\wp^\prime(z)$ is an odd function in $z$, $\wp^\prime(-z)=-\wp^\prime(z)$. Therefore, the integer $n$ needs to be adjusted according to the regime of the spiral when solving Eq.~\eqref{Eq:solution_weier}: $n=1$ for raising regime, and $n=-1$ for lowering regime.

\section{Physical discussion of the solutions}\label{Sec:connection}
Each family of solutions involves a fundamental curve: the case $\gamma=1$ relates to conic sections, $\gamma=2$ to logarithmic spirals, $\gamma=3$ to cardioids, and $\gamma=4$ to sinusoidal spirals. This section is devoted to analyzing the geometrical and dynamical connections between the solutions and other integrable systems.

%It is worth emphasizing that the solutions to the simple problem $\rho(\theta)$ transform to the solutions to Eq.~\eqref{Eq:two_body_problem}, $r(\theta)$, by means of the inverse transformation $\mathscr{S}^{-1}$. The transformation involves the free parameter $\xi$ and 

\subsection{Connection with Schwarzschild geodesics}

The Schwarzschild metric is a solution to the Einstein field equations of the form
\begin{equation}
	(\mathrm{d}s)^2 = \left( 1 - \frac{2M}{r} \right)\,(\mathrm{d}t)^2 - \frac{(\mathrm{d}r)^2}{1-2M/r} - r^2(\mathrm{d}\phi)^2 - r^2 \sin^2\phi\,(\mathrm{d}\theta)^2,
\end{equation}
written in natural units so that the speed of light and the gravitational constant equal unity, and the Schwarzschild radius reduces to $2M$. In this equation $M$ is the mass of the central body, $\phi=\piup/2$ is the colatitude, and $\theta$ is the longitude. 

The time-like geodesics are governed by the differential equation
\begin{equation}
	\left(\oder{r}{\theta}\right)^2 = \frac{1}{L^2}(E^2-1)\,r^4 + \frac{2M}{L^2}\,r^3 - r^2 + 2Mr ,
\end{equation}
where $L$ is the angular momentum and $E$ is a constant of motion related to the energy and defined by
\begin{equation}
	E = \left( 1-\frac{2M}{r} \right)\oder{t_s}{t_p} 
\end{equation}
in terms of the proper time of the particle $t_p$ and the Schwarzschild time $t_s$. Its solution is given by
\begin{equation}
	\theta(r) - \theta_0 = \int_{r_0}^r \frac{nL\,\mathrm{d}s}{\sqrt{f(s)}}
\end{equation}
and involves the integer $n=\pm1$, which gives the raising/lowering regime of the solution. Here $f(s)$ is a quartic function defined like in Eq.~\eqref{Eq:quartic_fs}. The previous equation is formally equivalent to Eq.~\eqref{Eq:integral_weier}. As a result, the Schwarzschild geodesics abide by Eqs.~\eqref{Eq:solution_weier} or~\eqref{Eq:solution_weier_symmetric}, which are written in terms of the Weierstrass elliptic functions and identifying the coefficients
\begin{equation}
\begin{split}
	&a_0 = E^2-1,\quad   && a_1=M/2,\quad  && a_2 =-L^2/6,\\
	&a_3 = ML^2/2,\quad  && a_4=0,\quad    && k   = nL.
\end{split}
\end{equation}
Compact forms of the Schwarzschild geodesics using the Weierstrass elliptic functions can be found in the literature \citep[see for example][\S4]{hagihara1930theory}. We refer to classical books like the one by \citet[][Chap.~3]{chandrasekhar1983mathematical} for an analysis of the structure of the solutions in Schwarzschild metric.

\begin{figure} 
	\centering
	\subfloat[$\gamma=3$\label{Fig:schwarz_g3}]{\includegraphics[width=.495\linewidth]{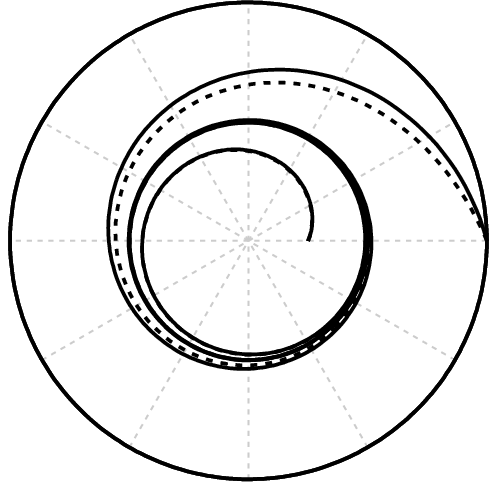}}
	\subfloat[$\gamma=4$\label{Fig:schwarz_g4}]{\includegraphics[width=.495\linewidth]{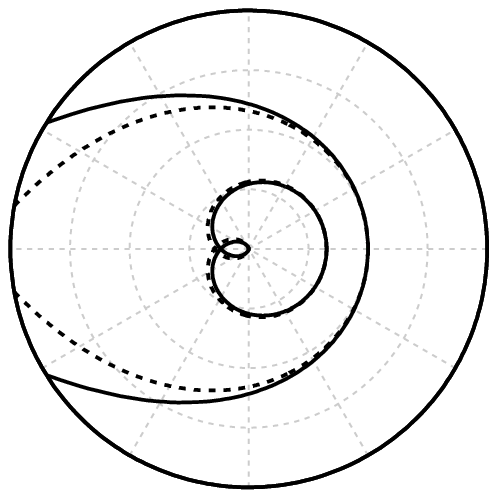}}
	\caption{Schwarzschild geodesic (solid line) compared to generalized cardioids and generalized sinusoidal spirals (dashed line).}
	% $M=3/14$ $E =0.9562$ $L= 0.7746$
\end{figure}

The analytic solution to Schwarzschild geodesics involves elliptic functions, like generalized cardioids ($\gamma=3$) and generalized sinusoidal spirals ($\gamma=4$). If the values of $\tka$ and $\tkb$ are adjusted in order the roots of $P_\mathrm{sol}$ to coincide with the roots of $f(r)$ in Schwarzschild metric, the solutions will be comparable. For example, when the polynomial $f(r)$ has three real roots and a repeated one the geodesics spiral toward or away from the limit radius
\begin{equation}
	r_\mathrm{lim} = \frac{1}{2M} - \frac{L^2+4M^2}{2ML(2L+\sqrt{L^2-12M^2})}.
\end{equation}
If we make this radius coincide with the limit radius of a hyperbolic cardioid with $\tkb=\sqrt{3\tka}$ [Eq.~\eqref{Eq:lim_radius_cardioid_Trans}] it follows
\begin{equation}
	\tka = \frac{1}{3r_\mathrm{lim}} \qquad \text{and} \qquad \tkb = \frac{1}{\sqrt{r_\mathrm{lim}}}.
\end{equation}
Figure~\ref{Fig:schwarz_g3} compares hyperbolic generalized cardioids and Schwarzschild geodesics with the same limit radius. The interior solutions coincide almost exactly, whereas exterior solutions separate slightly. Similarly, when $f(r)$ admits three real and distinct roots the geodesics are comparable to interior and exterior hyperbolic solutions of Type II. For example, Fig.~\ref{Fig:schwarz_g4} shows interior and exterior hyperbolic sinusoidal spirals of Type II with the inner and outer radii equal to the limit radii of Schwarzschild geodesics. Like in the previous example the interior solutions coincide, while the exterior solutions separate in time. In order for two completely different forces to yield a similar trajectory it suffices that the differential equation $\mathrm{d}r/\mathrm{d}\theta$ takes the same form. Even if the trajectory is similar the integrals of motion might not be comparable, because the angular velocities may be different. As a result, the velocity along the orbit and the time of flight between two points will be different. Equation~\eqref{Eq:to_integrate} shows that the radial motion is governed by the evolution of the flight-direction angle. For $\gamma=3$ and $\gamma=4$ the acceleration in Eq.~\eqref{Eq:ap_transformed} makes the radius to evolve with the polar angle just like in the Schwarzschild metric. Consequently, the orbits may be comparable. However, since the integrals of motion~\eqref{Eq:integr1} and~\eqref{Eq:integr2} do not hold along the geodesics, the velocities will not necessarily match.

%The similarity of the solutions is not by chance. The perturbation in Eq.~\eqref{Eq:ap_original} preserves an angular like-

\subsection{Newton's theorem of revolving orbits}\label{Sec:newton}
Newton found that if the angular velocity of a particle following a Keplerian orbit is multiplied by a constant factor $k$, it is then possible to describe the dynamics by superposing a central force depending on an inverse cubic power of the radius. The additional perturbing terms depend only on the angular momentum of the original orbit and the value of $k$.

Consider a Keplerian orbit defined as
\begin{equation}\label{Eq:kepler_newton}
	{\rho}(\theta) = \frac{\tilde{h}_1^2}{1+\tilde{e}\cos\nu_1}.
\end{equation}
Here $\nu_1=\theta-\theta_m$ denotes the true anomaly. The radial motion of hyperbolic spirals of Type II ---Eq.~\eqref{Eq:traj_hypII}--- resembles a Keplerian hyperbola. The difference between both orbits comes from the angular motion, because they revolve with different angular velocities. Indeed, recovering Eq.~\eqref{Eq:traj_hypII}, identifying $\tkb=\tilde{e}$ and ${\rho}_\mathrm{min}(1+\tkb)=\tilde{h}_1^2$, and calling the spiral anomaly $\nu_2=k(\theta-\theta_m)$, it is
\begin{equation}
	\rho(\theta) = \frac{\tilde{h}_1^2}{1+\tilde{e}\cos\nu_2}.
\end{equation}
When equated to Eq.~\eqref{Eq:kepler_newton}, it follows a relation between the spiral ($\nu_2$) and the true ($\nu_1$) anomalies:
\begin{equation}
	\nu_2 = k\nu_1.
\end{equation}
The factor $k$ reads
\begin{equation}
	k = \frac{\tkb}{\ell}=\frac{\tkb}{\sqrt{\tkb^2-1}}.
\end{equation}
Replacing $\nu_1$ by $\nu_2/k$ in Eq.~\eqref{Eq:kepler_newton} and introducing the result in the equations of motion in polar coordinates gives rise to the radial acceleration that renders a  hyperbolic generalized logarithmic spiral of Type II, namely
\begin{equation}
	\tilde{a}_{r,2} = -\frac{1}{{\rho}^2} + \frac{\tilde{h}^2_1}{{\rho}^3}(1-k^2) = \tilde{a}_{r,1} + \frac{\tilde{h}^2_1}{{\rho}^3}(1-k^2).
\end{equation}
This is in fact the same result predicted by Newton's theorem of revolving orbits.
  
The radial acceleration $\tilde{a}_{r,2}$ yields a  hyperbolic generalized logarithmic spiral of Type II. This is a central force which preserves the angular momentum, but not the integral of motion in Eq.~\eqref{Eq:integr2}. Thus, a particle accelerated by $\tilde{a}_{r,2}$ describes the same trajectory as a particle accelerated by the perturbation in Eq.~\eqref{Eq:ap_transformed}, but with different velocities. As a consequence, the times between two given points are also different. The acceleration derives from the specific potential 
\begin{equation}
	V({\rho}) = V_k({\rho}) + \Delta V({\rho}),
\end{equation}
where $V_k({\rho})$ denotes the Keplerian potential, and $\Delta V({\rho})$ is the perturbing potential:
\begin{equation}
	\Delta V({\rho}) = \frac{\tilde{h}_1^2}{2\rho^2}(1-k^2).
\end{equation}

\subsection{Geometrical and physical relations}
The inverse of a generic conic section $r(\theta)=a+b\cos\theta$, using one of its foci as the center of inversion, defines a lima\c{c}on. In particular, the inverse of a parabola results in a cardioid. Let us recover the equation of the trajectory of a generalized parabolic cardioid (a true cardioid) from Eq.~\eqref{Eq:traj_parab_gamma3}, 
\begin{equation}
	{\rho}(\theta) = \frac{{\rho}_\mathrm{max}}{2}[1+\cos(\theta-\theta_m)].
\end{equation}
Taking the cusp as the inversion center defines the inverse curve:
\begin{equation}
	\frac{1}{{\rho}} = \frac{2}{{\rho}_\mathrm{max}}[1+\cos(\theta-\theta_m)]^{-1}.
\end{equation}	
Identifying the terms in this equation with the elements of a parabola it follows that the inverse of a generalized parabolic cardioid with apoapsis ${\rho}_\mathrm{max}$ is a Keplerian parabola with periapsis $1/{\rho}_\mathrm{max}$. The axis of symmetry remains invariant under inversion; the lines of apses coincide.

The subfamily of elliptic generalized logarithmic spirals is a generalized form of Cotes's spirals, more specifically of Poinsot's spirals:
\begin{equation}
	\frac{1}{{\rho}} = a + b\cosh\nu.
\end{equation}
Cotes's spirals are known to be the solution to the motion immerse in a potential $V(r)=-\mu/(2r^2)$ \citep[][p.~69]{danby1992celestial}.

%Similarly, the radial motion ${\rho}(\theta)$ for Type II hyperbolic generalized logarithmic spirals is the same as a Keplerian orbit, the only difference being the angular motion. In Section~\ref{Sec:newton} it is proved that Newton's theorem of revolving orbits can be applied to determine an equivalent central force that renders the same exact trajectory, but with different velocities.

%The parabolic solutions for $\gamma>1$ share an interesting geometrical property: the evolute of the trajectory is a curve of the same type. That is, the evolute of a logarithmic spiral is a logarithmic spiral, just like the evolute of a cardioid is a cardioid and the 

The radial motion of interior Type II hyperbolic and Type I elliptic sinusoidal spirals has the same form, and is also equal to that of Type II hyperbolic generalized cardioids, except for the sorting of the terms.

It is worth noticing that the dynamics along hyperbolic sinusoidal spirals with $\tkb=2\tka$ are qualitatively similar to the motion under a central force decreasing with $r^{-5}$. Indeed, the orbits shown in Fig.~\ref{Fig:sinusoidal_trans} behave as the limit case $\gamma=1$ discussed by \citet[][Fig.~4]{macmillan1908motion}. On the other hand, parabolic sinusoidal spirals (off-center circles) are also the solution to the motion under a central force proportional to $r^{-5}$.

\section{Conclusions}

The dynamical symmetries in Kepler's problem hold under a special nonconservative perturbation: a disturbance that modifies the tangential and normal components of the gravitational acceleration in the intrinsic frame renders two integrals of motion, which are generalized forms of the equations of the energy and angular momentum. The existence of a similarity transformation that reduces the original problem to a system perturbed by a tangential uniparametric forcing simplifies the dynamics significantly, for the integrability of the system is evaluated in terms of one single parameter. The algebraic properties of the equations of motion dictate what values of the free parameter make the problem integrable in closed form.

The extended integrals of motion include the Keplerian ones as particular cases. The new conservation laws can be seen as generalizations of the original integrals. The new families of solutions are defined by fundamental curves in the zero-energy case, and there are geometric transformations that relate different orbits. The orbits can be unified by introducing the Weierstrass elliptic functions. This approach simplifies the modeling of the system.

The solutions derived in this paper are closely related to different physical problems. The fact that the magnitude of the acceleration decreases with $1/r^2$ makes it comparable with the perturbation due to the solar radiation pressure. Moreover, the inverse similarity transformation converts Keplerian orbits into the conic sections obtained when the solar radiation pressure is directed along the radial direction. The structure of the solutions, governed by the roots of a polynomial, is similar in nature to the Schwarzschild geodesics. This is because under the considered perturbation and in Schwarzschild metric the evolution of the radial distance takes the same form. The perturbation can also be seen as a control law for a continuous-thrust propulsion system. Some of the solutions are comparable with the orbits deriving from potentials depending on different powers of the radial distance. Although the trajectory may take the same form, the velocity will be different, in general.

\section*{Acknowledgements}
This work is part of the research project entitled ``Dynamical
Analysis, Advanced Orbit Propagation and Simulation
of Complex Space Systems'' (ESP2013-41634-P) supported
by the Spanish Ministry of Economy and Competitiveness. The author has been funded by a ``La Caixa'' Doctoral Fellowship for he is deeply grateful to Obra Social ``La Caixa''. The present paper would not have been possible without the selfless help, support, and advice from Prof. Jes\'us Pel\'aez. An anonymous reviewer made valuable contributions to the final version of the manuscript.

% Don't change these lines
\bsp	% typesetting comment
\label{lastpage}
\end{document}